\documentclass{aa}  
\usepackage{graphicx}
\usepackage{txfonts}
\usepackage{colortbl,array} % farbige cells
\usepackage{booktabs}
\usepackage[table]{xcolor}
\usepackage{fancybox}
\usepackage{natbib}

\bibpunct{(}{)}{;}{a}{}{,} % to follow the A&A style

\begin{document} 

\title{E-BOSS: An Extensive stellar BOw Shock Survey.}

\subtitle{II. Catalogue second release}

   \author{C. S. Peri\inst{1,2}\fnmsep\thanks{Fellow of CONICET},
           P. Benaglia\inst{1,2}\fnmsep\thanks{Member of CONICET},
           N. L. Isequilla\inst{2}}

   \institute{Instituto Argentino de Radioastronom\'{\i}a, 
     CCT-La Plata (CONICET), C.C.5, (1894) Villa Elisa, Argentina\\
     \email{cperi@fcaglp.unlp.edu.ar}
     \and
     Facultad de Ciencias Astron\'omicas y Geof\'isicas, UNLP, 
     Paseo del Bosque s/n, (1900) La Plata, Argentina}

  \date{Received July 25, 2014; accepted February 23, 2015}

% \abstract{}{}{}{}{} 
% 5 {} token are mandatory
 
  \abstract
   {Stellar bow shocks have been studied not only observationally,
but also theoretically since the late 1980s. Only a few catalogues
of them exist. The bow shocks show emission along all the 
electromagnetic spectrum, but they are detected more easily in 
infrared wavelengths. The release of new and high-quality infrared 
data eases the discovery and subsequent study of new objects.}
   {We search stellar bow-shock candidates associated with nearby 
runaway stars, and gather them together with those found elsewhere,
to enlarge the list of the E-BOSS first release. We aim to characterize 
the bow-shock candidates and provide a database suitable for statistical 
studies. We investigate the low-frequency radio emission at the position
of the bow-shock features, that can contribute to further studies of
high-energy emission from these objects.} 
   {We considered samples from different literature sources and searched 
for bow-shaped structures associated with stars in the Wide-field Infrared 
Survey Explorer (WISE) images. We looked for each bow-shock candidate 
on centimeter radio surveys.}
   {We reunited 45 bow-shock candidates and generated composed WISE images 
to show the emission in different infrared bands. Among them there are new 
sources, previously studied objects, and bow shocks found serendipitously. 
Five bow shocks show evidence of radio emission.}
   {Stellar bow shocks constitute an active field with open questions and 
enormous amounts of data to be analyzed. Future research at all wavelengths 
databases, and use of instruments like Gaia, will provide a more complete 
picture of these objects. For instance, infrared spectral energy distributions 
can give information about physical parameters of the bow shock matter.
In addition, dedicated high-sensitivity radio observations can help to 
understand the radio-$\gamma$ connection.}


   \maketitle
%
%________________________________________________________________

\section{Introduction}

High-mass stars interact with the interstellar medium (ISM)
producing different kinds of observable structures. An important 
proportion of early-type (OB) stars are runaway and have high 
velocities with respect to the normal stars, nearby ISM, and mean 
Galactic rotation speed (\citealt{Blaauw1961,Gies1986,Gies1987,
Stone1991,Tetzlaff2011,Fujii2011}). 

Some runaway stars generate the so-called stellar bow shocks,
a comma-shaped source usually detected in front of the star 
trajectory. The stellar motion with respect to the ISM helps 
to accumulate the matter ahead of the star, and the stellar 
ultraviolet radiation field heats the dust that emits infrared 
(IR) photons (e.g., \citealt{vanBuren1988,Noriega-Crespo1997,
Kobulnicky2010,Gvaramadze2014}). Additionally, the supersonic 
motion of a star produces shock waves, and through mechanisms 
like Fermi of first order, the charged particles can be 
accelerated up to relativistic velocities. These particles can 
interact with different fields --matter, magnetic, or photon 
fields-- and produce high energy and radio emission 
(\citealt{Benaglia2010,delValle2012,delValle2013,
delValle2014,Lopez-S2012}). The stellar UV field also ionizes 
the gas, which recombines and emits photons in the optical 
range (\citealt{Brown2005,Gull1979}). 

The infrared luminosity of stellar bow shocks is high compared 
to those in other spectral bands (e.g., \citealt[Fig.~5]{Benaglia2010}; 
\citealt[Fig.~11]{delValle2012}; \citealt[Table~1]{Benaglia2012}). 
Stellar bow shocks are in principle not difficult to discover at IR 
frequencies if they are close to us --no farther than a few kpc-- 
and not very dim.

Among the first results of the InfraRed Astronomical Satellite
(IRAS, \citealt{Neug1984}), the first list of stellar bow shocks 
was published in the late 1980s. \cite{Noriega-Crespo1997} found 
19 bow-shock candidates (BSCs, hereafter) and 2 doubtful objects 
among 58 high-mass runaway stars. The Wide-field Infrared Survey 
Explorer (WISE, \citealt{Wright2010}) satellite data release 
offered a sensitive arcsec resolution mid-infrared all-sky survey. 
In 2012, \cite{Peri2012} published the first release of the 
catalogue called E-BOSS: an Extensive stellar BOw Shock Survey 
(hereafter E-BOSS r1), done by searching these sources mainly in 
the then available WISE images (57\% of the sky). The final product 
was made up of a list of 28 BSCs, and minor studies of them in radio 
and optical wavelengths.

In this paper, we present the E-BOSS second release (hereafter 
E-BOSS r2), where we extended the search to more samples. The sources 
taken into account are described in Section 2; the different databases 
used to perform the search in Section 3; and the results, including 
the final BSCs release 2 list, in Section 4. Section 5 presents a 
discussion and Section 6 closes with a summary and conclusions.

%__________________________________________________________________

\section{Data samples: organization of the groups}

E-BOSS first release (r1, \citealt{Peri2012}) was based on two samples. 
Group 1 consisted of a set of sources taken from \cite{Noriega-Crespo1997} 
and Group 2 was a list extracted from \cite{Tetzlaff2011}. To build the 
present and second E-BOSS release (r2) we used samples extended to other 
spectral types, sources with no public WISE data at the moment of making 
E-BOSS r1, bow shocks from the literature, and bow shocks found 
serendipitously. In order to avoid confusion about the group numbers of 
r1 and r2, we begin here with Group 3. We give a first approximation for 
each group origin in this section; then we give a detailed analysis in the 
Results section.

\subsection{Group 3}

\cite{Tetzlaff2011} generated a catalogue of runaway stars using as the 
main source the {\it Hipparcos} stellar catalogue \citep{vanLeeuwen2007}.
The authors selected 7663 young stars up to 3 kpc from the Sun; completed 
the radial velocities from literature; studied peculiar spatial, radial, 
and tangential velocities; and estimated the runaway probability for each 
star of the sample. In addition, they investigated some stars that showed 
high velocities with respect to their environment. Previous works analyzed 
the runaway origin of the high-velocity stars taking into account only 
peculiar spatial velocities \citep{Blaauw1961,Gies1986}, radial velocities 
\citep{CruzGonzalez1974}, or tangential velocities \citep{Moffat1998}.
Instead, \cite{Tetzlaff2011} combined all the available criteria and 
give a list of 2547 runaway stars, increasing the amount of sources in 
two orders of magnitude.  

Originally, E-BOSS r1 Group 2 included 244 candidates for runaway stars 
from \cite{Tetzlaff2011}. The stellar spectral types range from O to B2, 
and 80 of them had no public WISE data at the moment of the first release. 
These 80 stars form Group 3 in this paper. We named the group Tetzlaff 
WISE 2, because the public IR data corresponds to the second WISE release. 
The list of Group 3 stars is shown in Table \ref{grupo3}.

\begin{table*}
\caption{Group 3: Tetzlaff WISE 2.}
\label{grupo3}
{\centering
{\small
\begin{tabular}{|ll|ll|ll|ll|ll|ll|}
\toprule
HIP                       & Sp.t.    & HIP    & Sp.t.    & HIP    & Sp.t.    & HIP    & Sp.t.   & HIP    & Sp.t.    & HIP   & Sp.t. \\
\midrule
3013                      & B2       & 38518  & B0.5Ib   & 39429  & O8Iaf    & 39776  & B2.5III & 40341  & B2V      & 41168 & B2IV  \\
41463                     & B2V      & 41878  & B1.5Ib   & 42316  & B1Ib     & 42354  & B2III   & 43158  & B0II/III & 43868 & B1Ib  \\
44251                     & B2.5V    &
\cellcolor{gray!50}44368  & B0.5Ib   & 46950  & B1.5IV   & 
\cellcolor{gray!50}47868  & B0IV     & 48469** & B1V     & 48527  & B2V      \\
48730                     & B2IV-V   & 48745  & B2III    & 49608  & B1III    & 49934  & B2IV    & 50899  & B0Iab/Ib & 51624 & B1Ib   \\
52526                     & B0Ib     & 52849  & O9V      & 52898  & B2III    & 54179  & B1Iab   & 54475  & O9II     & 58587 & B2IV   \\
61958                     & Op       & 65388  & B2       & 74368  & B0       & 89902  & B2V     & 94716  & B1II-III & 97045 & B0V    \\
97845                     & B0.5III  & 
\cellcolor{gray!50}98418  & O7       & 98661  & B1Iab    & 99283  & B0.5IV   & 99303  & B2.5V   & 99435  & B0.5V    \\ 
99580                     & O5e      & 99953  & B1V      & 100088 & B1.5V    & 100142 & B2V     & 100314 & B1.5Ia   & 100409 & B1Ib  \\
\cellcolor{gray!50}101186*& O9.5Ia   & 101350 & B0V      & 102999 & B0IV     & 103763 & B2V     & 104316 & O9       & 104548 & B1V   \\
\cellcolor{gray!50}104579 & B1V      & 104814 & B0.5V    & 
\cellcolor{gray!50}105186 & O8       & 105912 & B2II     & 106620 & B2V      & 106716 & B2V     \\
107864                    & Op       & 108911 & B2Iab    & 109051 & B2.5III  & 109082 & B2V     & 109311 & B1V      & 109332 & B2III  \\
109556                    & B1II     & 109562 & O9Ib     & 109996 & B1II     & 110025 & B2III   & 110287 & B1V      & 110362 & B0.5IV \\
110386                    & B2IV-V   & 110662 & B1.5IV-V & 110817 & B0.5Ib   & 111071** & B0IV  & 112482 & B1II     & 112698 & B1V    \\
114482                    & O9.5Iab  & 114685 & O7       &        &          &          &       &        &          &        &        \\
\bottomrule
\end{tabular}}
\tablefoot{Stars from E-BOSS r1 Group 2 without WISE observations: 80 sources
(identified by {\it Hipparcos} number). 
Spectral types (Sp. t.) range from O to B2, and were taken from \cite{Tetzlaff2011}.
Shaded in gray: 6 bow-shock candidates.
(*): analyzed on E-BOSS r1 using MSX data, 
(**): special cases, see Section 4.5.}}
\end{table*}

\subsection{Group 4}

Group 4 collects stars from \cite{Tetzlaff2011}, but with later 
spectral types with respect to Groups 2 and 3. We selected stars 
with spectral types from B3 to B5, and called this group Tetzlaff 
B3-B5. The list is in Table \ref{grupo4} and has a total of 234 
stars.

\begin{table*}
\caption{Group 4: Tetzlaff B3-B5.}
\label{grupo4}
{\centering
{\small
\begin{tabular}{| ll | ll | ll | ll | ll | ll |}
\toprule
HIP                      & Sp.t.      & HIP     & Sp.t.  & HIP     & Sp.t.   & HIP   & Sp.t.    & HIP     & Sp.t.  & HIP   & Sp.t.  \\
\midrule
   398                   & B3V        &   744   & B5V    &  1115   & B4V     &  1621 & B3       & 3478**  & B5V... &  3887 & B3Ia   \\
  4281                   & B5         &  4769   & B      &  4902   & B5      &  5023 & B        & 5062    & B3V    &  5569 & B5     \\
  6775                   & B3         &  7873   & B3V    &  9026   & B5      &  9549 & B5V      & 11487   & B5III  & 11607 & B5V    \\
 11894**                 & B3         & 12724   & B5     & 13187   & B3      & 14898 & B3V      & 15114** & B5Ve   & 15180 & B5III  \\ 
 15188                   & B3V        & 15424   & B5III  & 15535   & B3IV/V  & 15981 & B3III    & 16203   & B3III  & 16466 & B4V    \\  
\cellcolor{gray!50}17358 & B5III      & 17686   & B5     & 18871   & B3V     & 20860 & B5V      & 22075   & B5     & 24667 & B3     \\
 24674                   & B5III      & 24795   & B5     & 25066   & B3V     & 25235 & B3vw     & 25288   & B4IVn  & 25777 & B5     \\ 
 25906                   & B3II       & 25969   & B5     & 26602   & B4      & 26821 & B4/B5III & 27447   & B3II   & 27548 & B5     \\ 
 27683                   & B...       & 28949   & B5IV   & 28981   & B5      & 29213 & B4V      & 29681   & B5     & 29900 & B5IV/V \\ 
 30143                   & B3V        & 30169** & B5III  & 30943   & B5V     & 31068 & B3V      & 31642   & B5III  & 31875 & B3V    \\ 
 32220                   & B5         & 32269   & B5/B6V & 32786** & B5Iab/b & 32864 & B4IV     & 33490   & B3V    & 33509 & B5     \\ 
 33987                   & B5III      & 34485** & B5III  & 34611   & B5      & 35013 & B5V      & 35051   & B3Vn   & 35217 & B5III  \\ 
 35767                   & B4III      & 36024   & B5III  & 36040   & B5      & 36235 & B5       & 36246   & B5V    & 36323 & B5V    \\ 
 36682                   & B4/B5V     & 37245   & B3V    & 37345   & B4III   & 37444 & B4Iab    & 37524   & B4V    & 39184 & B5Vn   \\ 
 39776                   & B2/B3III   & 39866   & B3V    & 39943   & B4V     & 40430 & B+...    & 41599   & B3Vnne & 41823 & B3V    \\ 
 42038                   & B3V        & 42041   & B5V    & 42251   & B3ne    & 42605 & B3IV/V   & 43057   & B5Ib   & 43114 & B3V    \\ 
 43589                   & B3Vn       & 43878   & B5V    & 43955** & B3V     & 44105 & B5       & 44879   & B3IV/V & 45014 & B3III  \\ 
 45119                   & B4III      & 45145   & B5V    & 45372   & B5V     & 45563 & B3       & 45742   & B5V    & 45776 & B5III  \\ 
 45817                   & B5Vn       & 46224   & B4V    & 46296   & B3V     & 46329 & B5V      & 46470   & B5IV/V & 
\cellcolor{gray!50}46928 & B5        \\
 47005                   & B3/B4III   & 48440   & B3IV   & 48547  & B3/5V    & 48589** & B3V    & 48835   & B3V    & 49281 & B4:Vne \\ 
 50044                   & B4Ve       & 50519   & B5III  & 50764  & B5III    & 51940 & B5V:     & 52161   & B5Vn   & 53294 & B5III  \\ 
 53880                   & B5III      & 54082   & B3III  & 54226  & B+...    & 56709 & B5       & 57669   & B3V    & 57870 & B4III  \\ 
 59232                   & B3IV       & 59607   & B4III  & 60823  & B3V      & 61602 & Bp       & 62913** & B3Ib:  & 64622 & B4V:ne \\ 
 65020                   & B5III      & 66220   & B      & 66291  & B3p      & 66339 & B5e      & 66524   & B5II   & 67042 & B4V    \\ 
 68247                   & B4III      & 69122   & B5IV   & 69491  & B5V      & 69591 & B5V      & 69619   & B3p    & 69978 & B4IV/V \\ 
 70042                   & B3III/IV   & 73020   & B5V    & 74117  & B3V      & 74680 & B3V      & 74716   & B3IV   & 75959 & B3V    \\
 76416**                 & B5IV       & 78355   & B5IV   & 80405  & B4V      & 82596 & B4III    & 82617   & B3III  & 82658 & B5V    \\  
 82868                   & B3Vnpe     & 83629   & B5III  & 84260  & B3Vn     & 84282 & B4IV     & 85159   & B4IV   & 85357 & B3III  \\  
 85398                   & B5IV:      & 85919   & B5IV:  & 87280  & Bpsh     & 87886 & B5Vn     & 87928   & B4III  & 88156 & B      \\  
 88201                   & B3V        & 89061   & B3II   & 89956  & B4:Iae   & 89975 & B3V      & 90761   & B5     & 90992 & B5III  \\ 
 91713                   & B3IV/V     & 92038   & B5III  & 93396  & B5       & 93463 & B5/B6IV  & 93581   & B4Vn   & 93974 & B5     \\  
 94157                   & B5V        & 94385** & B3V    & 94391  & B4       & 94740 & B5       & 94859   & B5V    & 94899 & B3Vn   \\ 
 95372                   & B3IV       & 95624   & B5     & 95818  & B5Vn     & 95856 & B        & 95952   & B5III  & 96115 & B5     \\ 
 96254                   & B3III      & 97201   & B5     & 97611  & B5V      & 97680 & B3V      & 99349   & B3/B4IV & 
 99527$\dagger$          & B4Ieq      \\ 
 99618**                 & B5         & 100296  & B5     & 100308 & B        & 100392 & B5      & 100556  & B3II/III & 101112 & B5  \\ 
101634                   & B3         & 101909  & B3V    & 102943 & B5       & 104320 & B3V     & 104609  & B3       & 105164 & B5V \\ 
105268**                 & B3IVe      & 105690  & B5     & 
\cellcolor{gray!50}107789 & B5        & 108215  & B3IV   & 108597 & B5III    & 108975 & B3V     \\  
110298                   & B5IV       & 110603  & B5Iab  & 112790 & B5V      & 113577 & B5      & 114998  & B5II/III & 115186 & B3V   \\ 
115729                   & B3III      & 117100  & B      & 117290 & B5       & 117315 & B3V     & 117700  & B5       & 118214 & B4Vne \\ 
\bottomrule
\end{tabular}}
\tablefoot{Stars with spectral types from B3 to B5: 234 sources. 
The sources and spectral types were taken from \cite{Tetzlaff2011}.
($\dagger$): the star HIP 99527 has two possible spectral types, 
B4Ieq and K2Ib; we decided to select the first one.
Many selected stars have spectral type B, and are included here.
Shaded in gray there are the 3 bow-shock candidates. 
HIP 17358 was already studied on E-BOSS r1.
The sources marked with ** show no clear bow-shock 
structures, and we give more details in Section 4.5.}}
\end{table*}

\subsection{Group 5}

\cite{Maiz2004} produced an O galactic star catalogue, and 
classified 42 of them as runaways. We created Group 5, GOSC, 
with these 42 objects, and show the list in Table \ref{grupo5}. 
There are two subgroups in Group 5; the first corresponds to  
stars already analyzed in E-BOSS r1, and the second to stars 
studied in this paper.

\begin{table*}
\caption{Group 5: GOSC.}
\label{grupo5}
{\centering
{\small
\begin{tabular}{| ll | ll | ll | ll | ll | ll |}
\toprule
HIP    & Sp.t.   & HIP    & Sp.t.   & HIP    & Sp.t.      & HIP   & Sp.t. & HIP    & Sp.t.   & HIP    & Sp.t.   \\
\midrule
  1415 & O9III   &  11099 & ON8V    &  11473 & O9.5II-III & 18350 & O9.5  &  18614 & O7.5III &  22783 & O9.5Ia  \\ 
 24575 & O9.5V   &  27204 & O9.5V   &  28881 & O8 V       & 29147 & O7.5V &  39429 & O4I     &  43158 & O9.7Iab \\ 
 50253 & O9.5III &  52849 & O9V     &  69892 & O8         & 81377 & O9.5V &  84588 & O9.7Iab &  85331 & O6.5II  \\
 88469 & O7.5Iab &  93118 & O6.5III &  98530 & O9.5III    & 99580 & O5V   & 102999 & O9V     & 104316 & ON9V    \\    
105186 & O7.5III & 109556 & O6I     & 114482 & O9.5Iab    &       &       &        &         &        &         \\
\midrule
HD                     & Sp.t.      & HD       & Sp.t.   & HD     & Sp.t. & HD     & Sp.t.  & HD     & Sp.t. &  HD     & Sp.t. \\
\midrule
 12993                 & O6.5V      & 37043    & O9III   & 36879  & O7V     & 
\cellcolor{gray!50}57682 & O9I      & 60858    & O8V     & 105056 & ON9.7Ia \\   
105627                 & O9II-III   & 116852   & O9III   & 148546 & O9Ia  & 153919 & O6.5Ia & 163758 & O6.5Ia & 168941 & O9.5II-III \\
175754                 & O8 II      & 188209   & O9.5Iab & 191423 & O9III &        &        &        &        &        &            \\ 
\bottomrule
\end{tabular}}
\tablefoot{Stars tagged as runaway in \cite{Maiz2004}: 42 stars. 
Top: sources already analyzed on previous groups (1, 2, and 3). 
Bottom: sources not analyzed on any other group. Shaded in gray: 
bow-shock candidate.}}
\end{table*}

\subsection{Group 6}

\cite{Hoogerwerf2001} have studied the dynamic origin of 56 runaway 
stars and 9 pulsars. Of the 56 stars, we eliminated those coincident 
with the rest of the groups (from 1 to 5), and find 10 new objects 
to analyze. The list is presented in Table \ref{grupo6}, labeled as 
Group 6, and named \cite{Hoogerwerf2001}.

\begin{table*}
\caption{Group 6: \cite{Hoogerwerf2001}.}
\label{grupo6}
{\centering
\begin{tabular}{| ll | ll | ll | ll | ll |}
\toprule
HIP                      & Sp.t.     &  HIP  & Sp.t.  & HIP      & Sp.t.  & HIP    & Sp.t. &  HIP   & Sp.t.  \\
\midrule
 3881                    & B5V+...   & 20330 & B5     & 38455    & B2.5V  &  48943 & B5V   & 
\cellcolor{gray!50}86768 & B1.5V     \\
92609                    & B2II-IIIe & 97774  & B2III & 102274** & B5     & 103206 & B5IV  & 105811 & B0Ib    \\
\bottomrule
\end{tabular}
\tablefoot{Stars extracted from \cite{Hoogerwerf2001}.
We removed sources contained in groups 1 to 5, 
and 10 objects remained. Spectral types from Simbad.
Shaded in gray: bow-shock candidate. Marked with **: special case.}}
\end{table*}

\subsection{Group 7}

We built Group 7 with BSCs already analyzed by colleagues (38 
objects) and BSCs that appeared serendipitously in the IR images 
during our searches (7 objects). This group is called 'Serendipity 
and literature'. In Table \ref{grupo7}, we enumerate the objects, 
list the identification in the original work (whenever possible), 
and characterize the WISE emission. The first column shows the 
name we gave to the sources in this group. We used the prefix G7 
for each object and added a number that goes from 01 to 45. The 
last column has a comment about the region, or important information 
from the cited papers or added by us.

For Groups from 2 to 6 we selected runaway stars and searched for 
BSCs around them. In  Group 7 (and Group 1, from r1) this criterion 
was not valid; the BSCs structures were discovered directly by us 
or by other authors.

\begin{table*}
\caption{Group 7: Serendipity and literature.}
\label{grupo7}
{\centering
\begin{tabular}{| l l l l c c l l | }
\toprule
Number    & Name       & RA (J2000)  & DEC (J2000)  & Ref.& WISE        & Figure          & Comments                              \\
\midrule 
G7-01     & M17-S1     & 18:20:22.72 & -16:08:34.27 & A & $\bullet$     & --              & M17 region                            \\
G7-02     & M17-S2     & 18:20:25.88 & -16:08:32.48 & A & $\bullet$     & --              & M17 region                            \\
G7-03     & M17-S3     & 18:20:26.63 & -16:07:08.55 & A & $\bullet$     & --              & M17 region                            \\
G7-04     & RCW 49-S1  & 10:22:23.06 & -57:44:27.92 & A & $\supset$     & Fig. \ref{BSC3} & RCW 49 region                         \\
G7-05     & RCW 49-S2  & 10:24:03.12 & -57:48:36.00 & A & $\bullet$     & --              & RCW 49 region                         \\
G7-06     & RCW 49-S3  & 10:24:39.18 & -57:45:20.97 & A & $\bullet$     & --              & RCW 49 region                         \\
\midrule
G7-07     & K1         & 20:34:28.9  & +41:56:17.0  & B & $\supset$     & Fig. \ref{BSC3} & Cygnus-X region                       \\
G7-08     & K2         & 20:34:34.5  & +41:58:29.3  & B & $\supset$     & Fig. \ref{BSC3} & Cygnus-X region                       \\
G7-09     & K3         & 20:28:30.2  & +42:00:35.2  & B & $\supset$     & Fig. \ref{BSC3} & Cygnus-X region                       \\
G7-10     & K4         & 20:28:39.4  & +40:56:51.0  & B & $\supset$     & Fig. \ref{BSC3} & Cygnus-X region                       \\
G7-11     & K5         & 20:34:55.1  & +40:34:44.0  & B & $\supset$     & Fig. \ref{BSC3} & Cygnus-X region                       \\
G7-12     & K6         & 20:36:13.3  & +41:34:26.1  & B & $\supset$     & Fig. \ref{BSC4} & Cygnus-X region                       \\
G7-13     & K7         & 20:36:04.4  & +40:56:13.0  & B & $\supset$     & Fig. \ref{BSC4} & Cygnus-X region                       \\
G7-14**   & K8         & 20:20:11.6  & +39:45:30.1  & B & ?             & --              & Cygnus-X region, Special case         \\
G7-15**   & K9         & 20:25:43.9  & +38:11:13.2  & B & ?             & --              & Cygnus-X region, Special case         \\
G7-16     & K10        & 20:29:22.1  & +37:55:44.3  & B & $\supset$     & Fig. \ref{BSC4} & Cygnus-X region                       \\
G7-17*    & K11        & 20:33:36.1  & +43:59:07.4  & B & $\supset$     & Fig. \ref{BSC4} & Cygnus-X region, BD +43$^\circ$ 3654   \\
\midrule
G7-18     & G1         & 17:27:11.23 & -34:14:34.9  & C & $\supset$     & Fig. \ref{BSC4} & NGC 6357 region                       \\
G7-19     & G2         & 17:22:03.43 & -34:14:24.1  & C & $\supset$     & Fig. \ref{BSC4} & NGC 6357 region                       \\
G7-20**   & G3         & 17:28:21.67 & -34:32:30.3  & C & $\supset$     & Fig. \ref{BSC5} & NGC 6357 region, HD 319881            \\
G7-21**   & G4         & 17:18:15.40 & -34:00:06.1  & C & $\supset$     & Fig. \ref{BSC5} & NGC 6357 region, New RGB figure       \\
G7-22     & G5         & 17:22:05.62 & -35:39:55.5  & C & $\supset$     & Fig. \ref{BSC5} & NGC 6357 region, [N78] 34             \\
G7-23**   & G6         & 17:22:50.02 & -34:03:22.4  & C & $\supset$     & Fig. \ref{BSC5} & NGC 6357 region                       \\
G7-24     & G7         & 17:27:12.53 & -33:30:40.0  & C & $\supset$     & Fig. \ref{BSC5} & NGC 6357 region                       \\
G7-25     & G8         & 17:24:05.62 & -34:07:09.5  & C & $\supset$     & Fig. \ref{BSC5} & NGC 6357 region                       \\
\midrule
G7-26**   & HD 192281  & 20:12:33.12 & +40:16:05.45 & D & ?             & --              & Special case                          \\
\midrule
G7-27*    & Vela X-1   & 09:02:06.86 & -40:33:16.9  & E, F & $\supset$  & Fig. \ref{BSC1} & HIP 44368, Group 3, HMXB              \\
\midrule
G7-28     & 4U 1907+09 & 19:09:37.9  & +09:49:49    & F & $\supset$     & Fig. \ref{BSC6} & HMXB                                  \\
\midrule
G7-29     & 4U 1700-37 & 17:03:56.77 & -37:50:38.92 & G & ?             & --              & HMXB                                  \\
\midrule
G7-30     & J1117-6120 & 11:17:12.93 & -61:20:08.6  & H & $\supset$     & Fig. \ref{BSC6} & NGC 3603 region                       \\
\midrule
G7-31     & TYC 3159-6-1 & 20:18:40.37 & 41:32:45   & I & $\supset$     & --              & Cygnus-X region                       \\
\midrule
G7-32     & BD -14 5040 & 18:25:38.9  &-14:45:05.74 &J & $\supset$      & Fig. \ref{BSC6} & NGC 6611 region                       \\
G7-33*    & HD 165319   & 18:05:58.84 &-14:11:52.9  &J & $\supset$      & --              & NGC 6611 region, HIP 88652, Group 2   \\
G7-34     & Star 1      & 18:15:23.97 &-13:19:35.8  &J & $\supset$      & Fig. \ref{BSC6} & NGC 6611 region                       \\
\midrule
G7-35     & H1          & 15:00:58.55 & -63:16:54.7  & K & $\dagger$     & --              & Near: YSO and HH 139                  \\
G7-36     & H2          & 20:21:18.99 & +34:57:50.96 & L & $\dagger$     & --              & Probable YSO                          \\  
G7-37     & H3          & 20:34:12.92 & +41:08:15.94 & M & $\dagger$     & --              & EGGs (evaporating gaseous globules)   \\
G7-38**   & H4          & 05:46:51.51 & +25:03:48.18 & & $\supset$?      & --              & Probable BSC?                         \\
\midrule
G7-39     & SER1        & 08:58:29.4  & -43:25:09    &  & $\supset$      & Fig. \ref{BSC6} & Probably produced by TYC 7688-424-1   \\
G7-40     & SER2        & 10:03:42    & -58:30:28    &  & $\supset$      & Fig. \ref{BSC6} & Not clear                             \\
G7-41**   & SER3        & 10:38:19    & -58:53:22    &  & $\supset$      & Fig. \ref{BSC7} & Probably produced by HD 303197        \\
G7-42**   & SER4        & 23:46:37    & +66:46:20    &  & $\supset$      & Fig. \ref{BSC7} & Probably produced by HIP 117265       \\ 
G7-43     & SER5        & 07:06:33.6  & -11:17:24.5  &  & $\supset$      & Fig. \ref{BSC7} & Probably produced by HIP 34301        \\
G7-44     & SER6        & 01:11:24.3  & +57:33:38    &  & $\supset$      & Fig. \ref{BSC7} & Not clear                             \\
G7-45     & SER7        & 17:01:20    & -38:12:24.5  &  & $\supset$      & Fig. \ref{BSC7} & Not clear                             \\
\bottomrule
\end{tabular}
\tablefoot{Number: G7 stands for 'Group 7' and the number 
from 01 to 45 is specific of this table. 
For G7 01 to 34 we looked for bow-shock candidates in WISE 
images using sources from other works (references in column 5). 
G7 39 to 45 were serendipitously discovered during the catalogue 
production.
(*): already analyzed in other Groups (1 to 6) of E-BOSS r1 or r2. 
Name: extracted from the references (A to M) for G7 01 to 34.
Coordinates: for stars that produce the bow shock (whenever possible), 
or BSC apex coordinates estimated by us (visually), 
where the star could not be identified. 
$\supset$: BS shape observed in WISE images, 
$\supset$?: BSC to be confirmed, 
$\bullet$: emission excess in the region of BS,
?: doubtful cases, detailed in Section Special cases, 
$\dagger$: not enough resolution to identify bow-shock shape in WISE images. 
Figure: reference to the Figures in the present work.
Comments: a relevant comment for each case.
References in column 5; 
A: \cite{Povich2008}, B: \cite{Kobulnicky2010}, C: \cite{Gvaramadze2011b},
D: \cite{Arnal2011}, E: \cite{Kaper1997}, F: \cite{Gvaramadze2011a}, 
G: \cite{Ankay2001}, H: \cite{Gvaramadze2013}, I: \cite{Gvaramadze2014}, 
J:\cite{Gvaramadze2008}, K: \cite{Liu2011}, L: \cite{Magnier1999}, 
M: \cite{Sahai2012}.}}
\end{table*}

\subsection{Wolf-Rayet stars}

The \cite{Tetzlaff2011} catalogue includes a group of Wolf-Rayet 
stars in the analysis of possible runaway stars. The authors list 
as runaways the 18 stars WR 2, 11, 15, 24, 31, 46, 47, 52, 66, 79, 
110, 111, 121, 136, 138, 139, 153, and 156 (see \citealt{vanderHucht2001} 
for coordinates). We looked for IR bow-shaped emission at the WISE 
image cutouts and found no BSC related to the mentioned stars. The 
findings can be summarized as emission from discrete sources (WR 24, 
WR 153), high noise level (WR 111), wind bubble (WR 136 related to 
NGC 6888), and no emission above the noise for the rest of the 
runaway WR stars.

%__________________________________________________________________

\section{Searches}

Previous evidence showed that stellar bow shocks have high infrared 
luminosity (e.g., \citealt[Fig.~5]{Benaglia2010}; 
\citealt[Fig.~11]{delValle2012}; \citealt[Table~1]{Benaglia2012}) 
compared to those in other wavelengths, which is why we aimed the 
search at infrared frequencies. The WISE second release 
\citep{Wright2010} was published after E-BOSS r1 and completed 
the all-sky survey, allowing us to continue with BSCs studies.

We made use of the InfraRed Science Archive 
service\footnote{http://irsa.ipac.caltech.edu/}
(IRSA/IPAC NASA), and performed the search seeking a bow-shock shape 
in front of each selected runaway star or BSC already observed, for 
all the r2 Groups (3 to 7). We looked for the BS structure for each 
source in WISE four bands, and used the FITS images already processed. 
The extent of the fields was one square degree. At least two authors 
inspected the results.

We show a RGB (red-green-blue) composed image of the resulting BSCs,
except those cases where the sources had an associated WISE image 
already published. The RGB image shows the emission of the WISE bands 
1, 3, and 4 (3.4 $\mu$m, 12.1 $\mu$m, and 22 $\mu$m, respectively). 
In particular, WISE band 4 is useful for detecting warm dust 
\citep{Wright2010}.

We investigated the final list of BSCs in radio wavelengths. The case 
of BD +43$^\circ$ 3654 (\citealt{Benaglia2010}) is the first BSC with 
synchrotron emission; we checked whether any of the new BSCs constituted 
a similar case. We looked for BSCs in the NRAO/VLA Sky 
Survey\footnote{http://www.cv.nrao.edu/nvss/postage.shtml} (NVSS, 
\citealt{Condon1998}), which returns images  of the 1.4 GHz frequency 
emission in several formats.

%__________________________________________________________________

\section{Results}

\subsection{Bow-shock candidates: Groups 3 to 6}

Group 3, Tetzlaff WISE 2 (Table \ref{grupo3}), has 80 objects. We 
found six BSC and two doubtful cases. Since the group consists of
runaway stars, we associated each BSC with a star name. The six 
cases are: HIP 44368, HIP 47868, HIP 98418, HIP 101186, HIP 104579, 
and HIP 105186, and are shaded in gray in Table \ref{grupo3}. The 
HIP 101186 region was already studied in E-BOSS r1, but with Midcourse 
Space Experiment (MSX, \citealt{Egan2003}) cutouts; no WISE data was 
available in the zone at that moment. For the six BSC we built RGB 
images from WISE bands cutouts (Figs. \ref{BSC1} and \ref{BSC2}). 
Cases like HIP 48469 and HIP 111071 have no clear bow-shock shapes 
in their surroundings; they are separated as special cases (Section 
4.5) and marked with ** in Table \ref{grupo3}.

Group 4 is also made up of runaway stars; hence, we use the star 
names to identify the BSCs. We found 3 BSCs and 13 special cases; 
they are shaded in gray and marked with **, respectively, in Table 
\ref{grupo4}. The BSCs are related to HIP 17358, HIP 46928, and 
HIP 107789. HIP 17358 was already analyzed in E-BOSS r1. Figure 
\ref{BSC2} shows HIP 46928 and 107789. Additionally, we encountered 
two bow-shock features not associated with the stars in Group 4, 
but in the stellar fields of HIP 5569 and HIP 33987. They will be 
presented in next section (Figs. \ref{BSC6} and \ref{BSC7}).

In Group 5 we found evidence of one BSC, identified around HD 57682. 
The source is shaded in gray in Table \ref{grupo5} and presented 
in Figure \ref{BSC2}. We found one BSC in Group 6 associated with 
HIP 86768 (Fig. \ref{BSC2}) and one special case around HIP 102274. 
Both cases are marked in Table \ref{grupo6}, in gray and with ** 
respectively.

\subsection{Bow-shock candidates: Group 7}

Group 7 BSCs are shown in Table \ref{grupo7}. \cite{Povich2008} 
reported the discovery of six infrared stellar wind bow shocks 
in the galactic massive star-forming regions M17 and RCW 49 from 
{\it Spitzer} GLIMPSE (Galactic Legacy Infrared Mid-Plane Survey 
Extraordinaire, \citealt{Benjamin2003}) images. We searched for 
them in the WISE image database. The images looked saturated,
except for the BSC RCW 49 S1 (Fig. \ref{BSC3}).

\cite{Kobulnicky2010} found ten stellar bow shocks among {\it Spitzer} 
images, in the Cygnus OB2 association, at the heart of the Cygnus-X 
region. An extra case is the star BD +43$^\circ$ 3654, studied by 
\cite{Comeron2007}, for which the authors confirm its runaway nature,
but display no image. We found each BSC in WISE images, but BSCs 
numbers 8 and 9 are doubtful cases (Section 4.5). We show WISE 
images in Figures \ref{BSC3} and \ref{BSC4}, and add the BSCs in 
the final E-BOSS r2 list. The WISE first release did not cover the 
region of BD +43$^\circ$ 3654, and we show the RGB WISE composite 
image in this work (Fig. \ref{BSC4}).

The next eight BSCs are located near two young clusters associated 
with the star-forming region NGC 6357 \citep{Gvaramadze2011b}. 
Seven were discovered in MSX, {\it Spitzer}, and WISE images. The 
authors presented one BSC with WISE images; hence, we constructed 
images for the other seven (Figs. \ref{BSC4} and \ref{BSC5}) and 
included all eight in the E-BOSS r2 final list.

The structure and kinematics of the ISM around the runaway star HD 
192281 was studied by \cite{Arnal2011}. The authors found signs of 
the star trajectory using Canadian Galactic Plane Survey (CGPS, 
\citealt{Taylor2003}) and MSX images. We looked for similar structures
in WISE images but no clear feature was found. In addition, we looked 
for a comma-shaped or similar emission, but with no success.

Vela X-1 (HIP 44368, Group 3) is the first case of a runaway HMXB 
(high-mass X-ray binary) showing a bow-shock structure ahead of its 
motion direction \citep{Kaper1997}. We include this source on E-BOSS r2. 
The star 4U 1907+09 is also a runaway object (\citealt{Gvaramadze2011a})
and presents a bow-shock feature (Fig. \ref{BSC6}). This is the second 
BSC associated with a high-mass X-ray binary. 4U 1700-37 is a runaway 
high-mass X-ray binary too (\citealt{Ankay2001}). We did not find a 
clear comma-shaped structure ahead of the star in any of the WISE bands.

\cite{Gvaramadze2013} studied two massive stars possibly ejected from 
the NGC 3603 cluster through dynamical few-body encounters. The authors 
showed {\it Spitzer} images of the surroundings of the star J1117-6120 
and its suspected original companion (a O2 If*/WN6 star, WR42e). We 
constructed a RGB WISE image for J1117-6120 (Figure \ref{BSC6}); for 
WR42e, the field is saturated.

Optical spectroscopy for the runaway blue supergiant TYC 3159-6-1 star 
was presented by \cite{Gvaramadze2014}, together with a discussion that 
assumes Dolidze 7 to be its parent cluster (Cygnus-X region). The authors 
show a band 3 (12 $\mu$m) WISE image. We include the TYC 3159-6-1 nebulae 
as a BSC (Table \ref{grupo7}).

Three BSCs were analyzed by \cite{Gvaramadze2008} near the NGC 6611 
cluster, located in the Ser OB 1 association. NGC 6611 has possibly 
ejected three of its O massive stars: BD -14$^{\circ}$ 5040, HD 165319, 
and `star 1'. HD 165319 (HIP 88652) was part of E-BOSS r1. For BD 
-14$^{\circ}$ 5040 and `star 1', we generated RGB WISE images (Fig. 
\ref{BSC6}) and included them in the E-BOSS r2 list (Table \ref{tablafinal}).

Sources G7 35 to 38 (Table \ref{grupo7}) were shown in a Hubblesite 
News Center 
release\footnote{http://hubblesite.org/newscenter/archive/releases/star/bow-shock/2009/03/}.
We named the objects H1, H2, H3, and H4. Observations made by R. Sahai 
(HST Cycle 14 proposal 10536), using the ACS SNAPshot Survey of the 
HST between 2005 and 2006, were dedicated to a survey of preplanetary 
nebulae candidates (\citealt{Sahai2007}). Among the observed sources, 
Sahai and his team found what seemed to be four BSC.

\cite{Liu2011} developed a thorough study of the young cluster associated 
with the Circinius molecular cloud, and uncovered a population of YSOs 
in the western region of the cloud. The authors found two aggregates of 
YSOs, and H1 is located in the middle region between both YSO groups. 
Several sources from different works and wavelengths are mixed inside 
the WISE source associated with H1, like the Herbig-Haro object 139. 
Although HH 139 is embedded in the WISE source, H1 is about 4 arc seconds 
from HH 139 at optical wavelengths. We do not have sufficient evidence 
yet to confirm the stellar bow-shock nature of H1.

Source H2 seems to be related to the infrared source IRAS 20193+3448, 
its nearest source. This source was identified as `a likely transitional 
YSO' \citep{Magnier1999}. If H2 is related to IRAS 20193+3448 and its 
surroundings, we rule out that it is a BSC, because the whole region is 
related to a star-forming region.

\cite{Sahai2012} published a multi-wavelength study of the source IRAS 
20324+4057 and some surrounding structures; all of them near H3. They 
conclude the sources were dense molecular cores originated in the Cygnus 
cloud. Therefore, we excluded H3 as a possible BSC like the ones we are 
looking for.

In Hubble images H4 has a boomerang shape, but the WISE source covers 
several times the optical one. There is no catalogued source around H4 
on about 30 arcseconds; therefore, we cannot discard the boomerang as 
a BSC.

Hubble Space Telescope has an angular resolution of 0.05 arcseconds, 
and WISE between 6.1 and 12 arcseconds depending on the band. For all 
the cases (H1, H2, H3, and H4), the IR sources are unresolved. Several 
optical sources can contribute to the IR emission. Only H4 remains as 
a possible BSC.

The last seven cases in Table \ref{grupo7} (G7 39 to 45) are serendipity 
BSCs. We found them in WISE images during the searches. We called the 
objects SER 1 to 7. SER 1 (Fig. \ref{BSC6}) seems to have been generated 
by the TYC 7688-424-1 star according to Simbad sources (spectral type 
B5Ve). For SER 2 (Fig. \ref{BSC6}) we do not find any catalogued star 
to be considered as a bow-shock producer. SER 3 (Fig. \ref{BSC7}) could 
be related to HD 303197, according to Simbad. HD 303197 (spectral type 
B5) is located near the NGC 3324 open cluster, although the bow-shock 
shape seems to indicate that the star might be coming from another 
direction, towards the cluster. SER 4 (Fig. \ref{BSC7}) could be produced 
by HIP 117265 (spectral type B2IV). SER 5 (Fig. \ref{BSC7}) can be 
generated by HIP 34301 (TYC 5389-3064-1), or BD-11 1790C (double system). 
For SER 6 (Fig. \ref{BSC7}) we did not find a star that might produce 
the BSC. SER 7 (Fig. \ref{BSC7}) has no clear star related to it; the 
nearest stars are HD 153426 (80.6''), and HD 322486 (102.2''), and they 
are not located where runaway stars usually lie to cause a bow-shock 
structure.

\subsection{Bow-shock candidates list}

In Table \ref{tablafinal} we list the final set of new BSCs that 
we found, together with those already analyzed by other authors. 
In E-BOSS release 2 there are 45 new objects in addition to those 
of release 1. The first column of Table \ref{tablafinal} gives 
the name of the star that generates the BSC whenever identification 
was possible, or another name when the star cannot be determined. 
The second column contains the Group number and the third column 
the galactic coordinates (of stars or stagnation point), from Simbad 
or reference from Table 5. The stellar spectral types were taken 
from Simbad (B-type), the GOS Catalog \citep{Maiz2004}, or the 
reference from Table 5. The distances were taken from \cite{Megier2009}, 
\cite{Mason1998} or were estimated using parallaxes from {\it Hipparcos} 
\citep{vanLeeuwen2007}. The stellar wind terminal velocities were 
interpolated or extrapolated from Table 3, \cite{Prinja1990}, or 
adopted as in \cite{Peri2012} for stars with the same spectral types.
The mass-loss rates were interpolated or extrapolated from \cite{Vink2001}.
The tangential velocity of HIP 44368 was estimated through proper 
motion values, and for the other sources of Groups 3 and 4 the 
values are from \cite{Tetzlaff2011}. Radial velocities for Groups 
3 to 6 are from \cite{Kharchenko2007}; for objects in Group 7, 
references are as in Table 5, and Simbad for SER 1, 3, 4, and 5.
Proper motions of stars in Groups 3 to 6 and K3 (HD 195229) are 
from {\it Hipparcos} \citep{vanLeeuwen2007}, and for G1 to G8 we 
show two values from \cite{Gvaramadze2011b}; for TYC 3159-6-1, 
BD -14 5040, and Star 1, references are as in Table 5 (see errors 
for velocities and proper motions in references); for SER 1, 3, 4, 
and 5, Simbad.

\begin{table*}
\caption{List of the E-BOSS release 2 bow-shock candidates and corresponding stellar parameters.}
\label{tablafinal}
{\small
\centering
%{\small
\begin{tabular}{| l@{~~~}c@{~~~}r@{~~~}r@{~~~}l@{~~~}c@{~~~}c@{~~~}c@{~~~}c@{~~~}c@{~~~}c@{~~~}c |}
\toprule
Source       & Gr.& $l$    & $b$    & Spectral type   & $d$            & $v_{\infty}$ & $\dot{M} \times 10^{6}$ 
& $v_{\rm tg}$ & $v_{\rm r}$   & $\mu_{\alpha} \cos \delta$ & $\mu_{\delta}$  \\
             &   & [\degr] & [\degr]  &               & [pc]           & [km/s]   & [$M_{\odot}$/yr] & [km/s] & [km/s] & [mas/yr] & [mas/yr] \\
\midrule
HIP 44368$^1$ &3,7& 263.1  & +3.9   & B0.5 Iab        &  1900$\pm$0.1$^z$ & 1100   & 0.8  & 52.2 & -1.00        &  -5.5     &   8.8     \\
HIP 46928     & 4 & 295.6  & -21.04 & B5V             &  [175.44]         & 100   & 0.03 & 13.6 & -42.00       & -34.81    &  14.18    \\
HIP 47868     & 3 & 261.8  & +17.4  & B0.5IIIn        &  [1075.27]        & 1200  & 0.3  & 29.5 & 31.70        & -11.44    &   5.92    \\
HIP 98418     & 3 & 71.6   & +2.9   & O7              &  [529.10]         & 2545  & 0.24 & 21.8 &  20.00       &  -5.56    &  -9.59    \\
HIP 104579    & 3 & 81.0   & -8.07  & B1Vp            &  [1149.42]        & 650   & 0.03 & 26.6 & -6.00        &   0.37    &   0.53    \\
HIP 105186    & 3 & 87.6   & -3.8   & O8e             &  1130$\pm$190$^a$ & 2340  & 0.1  & 57.0 & -30.00       &   4.85    &  -8.40    \\
HIP 107789    & 4 & 102.1  & +4.8   & B5              &  [1190.47]        & 100   & 0.03 & 16.3 & -16.00       &  -1.5 1   &  -5.07    \\
HD 57682     & 5 & 224.4   & +2.6   & O9 IV            & 1600$^b$         & 1900   & 0.16 &  --  & 24.10       &  10.46     &  13.38     \\
HIP 86768    & 6 & 18.7    & +11.6  & B1.5V            & 737$^a$          & [550]  & 0.03 &  --  & -26         & -4.32      & -10.60     \\
M17-S1       & 7 & 15.07   & 0.64   & O9-B2V           & 1600$^z$         & 1000   & 0.03  & --  & --          & --         & --         \\
M17-S2       & 7 & 15.08   & 0.65   & O7-O8V           & ''               & [1500] & 0.16  & --  & --          & --         & --         \\
M17-S3       & 7 & 15.10   & 0.64   & O7V              & ''               & 2300   & 0.25  & --  & --          & --         & --         \\
RCW 49-S1    & 7 & 284.08  & 0.43   & O5III            & 6100$^z$         & 2800   & 3.23  & --  & --          & --         & --         \\
RCW 49-S2    & 7 & 284.30  & 0.3    & O6 V             & ''               & 2600   & 0.6   & --  & --          & --         & --         \\
RCW 49-S3    & 7 & 284.34  & 0.2    & O3V/O6.5III      & ''               & 2800   & 2     & --  & --          & --         & --         \\
K1           & 7 & 80.86   & 0.97   & O9V              & 1500$^z$         & 1500   & 0.05  & --  &-17 $\pm$ 6  & --         & --         \\
K2           & 7 & 80.90   & 0.98   & B1V-B3V          & ''               &  500   & 0.03  & --  &-12 $\pm$ 15 & --         & --         \\
K3           & 7 & 80.26   & 1.91   & B0.2III          & ''               & [1250] & 0.1   &     &-3 $\pm$ 2   & -0.22      & 3.84       \\
K4           & 7 & 79.42   & 1.28   & B2V-B3V          & ''               & [300]  & 0.03  & --  & --          & --         & --         \\
K5           & 7 & 79.82   & 0.09   & O9V              & ''               & 1500   & 0.05  & --  & 10 $\pm$ 10 & --         & --         \\
K6           & 7 & 80.76   & 0.49   & B4V-B6V          & ''               & 250    & 0.03  & --  & --          & --         & --         \\
K7           & 7 & 80.24   & 0.14   & O5V              & ''               & 2500   & 1.5   & --  & --          & --         & --         \\
K8           & 7 & 77.52   & 1.90   & B1V-B3V          & ''               & [400]  & 0.03  & --  & 2 $\pm$ 4   & --         & --         \\
K9           & 7 & 76.84   & 0.12   & B?               & ''               & 400    & 0.03  & --  & --          & --         & --         \\
K10          & 7 & 77.05   &-0.61   & B1V-B2V          & ''               & [550]  & 0.03  & --  & --          & --         & --         \\
G1           & 7 & 353.42  & 0.45   & O7.5-O7V         & 1700$^z$         & 2100   & 0.2   & --  & --          & -1.5/-3.8  & -1.6/-5    \\
G2           & 7 & 352.82  & 1.33   & O5.5-O6.5-/V     &     ''           & 2250   & 0.4   & --  & --          & -8.5/-8.9  & -3.2/-11.3 \\
G3           & 7 & 353.30  & 0.08   & O6Vn-O5V         &     ''           & 2000   & 0.4   & --  & --          & 0/3.4      & -2.8/-3.2  \\
G4           & 7 & 352.57  & 2.11   & O6.5-O6V         &     ''           & 2550   & 0.5   & --  & --          & -4.4/-9    & 0.9/2.3    \\
G5           & 7 & 351.65  & 0.51   & O8/B0III/V-O6.5V &     ''           & 2000   & 0.1   & --  & --          & -4.9/-11.8 & 11.7/18.7  \\
G6           & 7 & 353.06  & 1.29   & B0V              &     ''           & [1000] & 0.1   & --  & --          & -4.6/-8.1  & 0.8/0.1    \\
G7           & 7 & 354.03  & 0.85   & B0V              &     ''           & [1000] & 0.1   & --  & --          & -3.9/-7.3  & 0.0/-3.2   \\
G8           & 7 & 353.16  & 1.05   & O9-9.5V          &     ''           & [1500] & 0.04  & --  & --          & -6.0/-9.4  & 1.9/-2.5   \\
4U 1907+09   & 7 & 43.74   &  0.47  & O9.5 Iab         & 4000$^z$         & 2900   & 0.7   & --  & --          & --         & --         \\
J1117-6120   & 7 & 291.88  & -0.50  & O6 V             & 7600$^z$         & 2600   & 0.6   & --  & -21.4       & --         & --         \\
TYC 3159-6-1 & 7 & 78.83   & +3.15  & O9.5-O9.7 Ib     & 1500$^z$         & 2900   & 0.7   & --  & -35.8       & -2.4       & -0.1       \\
BD -14 5040  & 7 & 16.89   & -1.12  & B                & 1800$^z$         & 400    & 0.03  &     & --          &  5.5/7.7   & -3.0/-4.6  \\
Star 1       & 7 & 16.98   &  1.75  & O9.5III/O5V--O7.5III/O4V & 1800$^z$ & 2200   & 0.63  &     & --          &  0/-4.3    & 12/0.9     \\
SER1         & 7 & 264.78  &  1.54  & B5 Ve            & --               & 250    & 0.03  & --  & --          & -9.5       & 8.5        \\
SER2         & 7 & 282.48  & -2.46  & --               & --               & --     & --    & --  & --          & --         & --         \\
SER3         & 7 & 286.46  & -0.34  & B5 (V)           & --               & 250    & 0.03  & --  & --          & --         & --         \\
SER4         & 7 & 116.59  &  4.70  & B2 IV            & --               & 500    & 0.03  & --  & -9.7        & 8.71       & -3.59      \\
SER5         & 7 & 224.69  & -1.82  & B0.5IV           & --               & 550    & 0.03  & --  & 31          & -3.14      & 3.32       \\
SER6         & 7 & 125.62  & -5.20  & --               & --               & --     & --    & --  & --          & --         & --         \\
SER7         & 7 & 347.15  &  2.36  & --               & --               & --     & --    & --  & --          & --         & --         \\
\bottomrule
\end{tabular}}
\tablefoot{Column 1: star names for the identified cases 
or name given by us; (1): Vela X-1. Column 2: group number 
for each object. 
Galactic coordinates l,b: from Simbad or corresponding 
references for Group 7 sources (Table 5, fifth column). 
Spectral types: for B-type stars from Simbad; 
for O-type stars (whenever possible) from GOS Catalog 
\citep{Maiz2004}, otherwise from same reference as in Table 5. 
For SER 2, 6, and 7 we did not find stars related.
Distances: (a) \cite{Megier2009}, (b) \cite{Mason1998}, 
(z) from same references of Table 5; 
[]: derived from {\it Hipparcos} parallaxes \citep{vanLeeuwen2007}. 
Wind terminal velocities: inter or extrapolated from Table 3, 
\cite{Prinja1990}; []: as adopted in \cite{Peri2012} for stars 
with same spectral types. 
Mass loss rates: inter or extrapolated from \cite{Vink2001}. 
Tangential velocities: for HIP 44368 estimated through proper motion values; 
for the other sources of groups 3 and 4 from \cite{Tetzlaff2011}. 
Radial velocities: for groups 3 to 6, from \cite{Kharchenko2007}; 
for objects from Group 7, references as in Table 5, 
and Simbad for SER 1, 3, 4, and 5.
Proper motions: for groups 3 to 6, and K3 (HD 195229),
from \cite{vanLeeuwen2007}; for G1 to G8 we show two values 
from \cite{Gvaramadze2011b}; for TYC 3159-6-1, BD -14 5040, 
and Star 1, reference as in Table 5 (see errors for velocities and
proper motions in references); for SER 1, 3, 4, and 5, Simbad.}
\end{table*}

\subsection{Bow-shock candidate features}

Table \ref{medidas} shows the measured sizes of the bow-shock 
candidates. We list the length $l$ and width $w$ of each 
bow-shock structure, and distance $R$ from the star to what 
is known as stagnation point (\citealt{Wilkin1996}), in angular 
units. The sizes were taken from WISE band 4 or band 3 images. 
Whenever possible, we also estimated $l$, $w$, and $R$ on linear 
units (pc) through distance measurements. No size is given for 
cases without WISE images.

The ISM ambient density was derived for the bow-shock candidates 
with adopted distances and stellar spectral types, as in 
\cite{Peri2012}. To this aim, we used the expression that gives 
the stagnation radius $R_0$: 

$$R_0 =  \sqrt{ \frac{ \dot{M} v_\infty }{ 4 \pi \rho_{\rm a} v_*^2} } \, ,$$

\noindent where $\rho_{\rm a} = \mu \, n_{\rm ISM}$ is the ambient 
medium density, and $v_*$ is the spatial stellar velocity. 
The volume density of the ISM is in H atoms at the bow-shock 
position assuming $R_0 \sim R$, a mass per H atom 
$\mu = 2.3 \times 10^{-24}$ g, and the helium fractional abundance 
$Y = 0.1$. For those stars with unknown spatial peculiar velocity  
we adopted $v_*$ = 25 km s$^{-1}$. We show the results in Table \ref{medidas}.
Values obtained for $n_{\rm ISM}$ must be taken into account with caution:
errors on each parameter used in the $R_0$ expression can have a strong 
influence in calculations.

\begin{table*}
\caption{Parameters of the bow-shock candidates and medium density.}
\label{medidas} 
{\small \centering
\begin{tabular}{|l@{~~~}|c@{~~~}c@{~~~}c@{~~~}|c@{~~~}c@{~~~}c@{~~~}|c@{~~~}||l@{~~~}|c@{~~~}c@{~~~}c@{~~~}|c@{~~~}c@{~~~}c@{~~~}|c@{~~~}|}
\toprule
Source              & $l$ & $w$  & $R$ &  $l$ & $w$ & $R$ & $n_{\rm ISM}$ & 
Source              & $l$ & $w$  & $R$ &  $l$ & $w$ & $R$ & $n_{\rm ISM}$ \\ 
                    &     & [']  &     &      & [pc]&     & [cm$^{-3}$]  &
                    &     & [']  &     &      & [pc]&     & [cm$^{-3}$]  \\
\midrule
HIP 44368           & 5    & 1.25 & 0.8 & 2.76 & 0.69 &	0.44 & 1.8  & G1          &  4  & 0.8  & 0.8 & 1.98 & 0.4  & 0.4  & 14  \\
HIP 46928           & 1.8  & 0.47 & 1   & 0.09 & 0.02 & 0.05 & 2    & G2          &  6  & 1.2  & 1.2 & 2.97 & 0.59 & 0.59 & 14  \\
HIP 47868           & 5    & 1.7  & 1.6 & 1.56 & 0.53 & 0.50 & 2.6  & G3          &  5  & 1.5  & 1   & 2.47 & 0.74 & 0.49 & 16  \\
HIP 98418           & 1.66 & 0.6  & 0.5 & 0.26 & 0.09 & 0.08 & 380  & G4          &  3  & 1    & 0.8 & 1.48 & 0.49 & 0.4  & 42  \\
HIP 101186*         & 22   & 4    & 4.5 & 9.50 & 1.73 & 1.94 & 2.63 & G5          &  3  & 1    & 1   & 1.48 & 0.49 & 0.49 & 4   \\  
HIP 104579          & 3    & 1.3  & 1.1 & 1.00 & 0.43 & 0.37 & 0.7  & G6          & 1.5 & 0.5  & 0.4 & 0.74 & 0.25 & 0.2  & 13  \\    
HIP 105186          & 7    & 2    & 2.5 & 2.30 & 0.66 & 0.82 & 0.3  & G7          & 1.5 & 0.5  & 0.3 & 0.74 & 0.25 & 0.15 & 23  \\
HIP 107789          & 1.2  & 0.4  & 0.3 & 0.42 & 0.14 & 0.10 & 1.8  & G8          & 1.5 & 0.5  & 0.5 & 0.74 & 0.25 & 0.25 & 5   \\
BD +43$^{\circ}$3654* & 19   & 3    & 4   & 8.01 & 1.26 & 1.69 & 3.5  & 4U 1907+09  & 3.5 & 1    & 1.5 & 4.07 & 1.16 & 1.74 & 0.1 \\  
HD 57682            & 1.6  & 0.4  & 0.3 & 0.74 & 0.19 & 0.14 & 85   & J1117-6120  & 1.5 & 0.5  & 0.5 & 3.31 & 1.1  & 1.1  & 6.4 \\
HIP 86768           &   4  & 2    &   2 & 0.86 & 0.43 & 0.43 & 0.1  & BD -14 5040 & 4   & 1    & 1   & 2.09 & 0.52 & 0.52 & 0.1 \\
RCW 49-S1           &  2.5 & 0.5  & 0.7 & 4.43 & 0.89 & 1.24 & 30   & Star 1      & 7   & 1    & 1.5 & 3.66 & 0.52 & 0.78 & --  \\
K1                  &   3  & 1    & 0.7 & 1.31 & 0.44 & 0.31 & 4    & SER1        & 4   & 1.5  & 1   & --   & --   & --   & --  \\
K2                  &   4  & 0.3  & 0.5 & 1.74 & 0.13 & 0.22 & 1.7  & SER2        & 2.5 & 1    & 1   & --   & --   & --   & --  \\
K3                  &   2  & 0.5  & 0.4 & 0.87 & 0.22 & 0.17 & 16.6 & SER3        & 2.5 & 0.5  & 0.5 & --   & --   & --   & --  \\
K4                  &  1   & 0.3  & 0.3 & 0.44 & 0.13 & 0.13 & 2.8  & SER4        & 2.5 & 1    & 1   & --   & --   & --   & --  \\
K5                  &    4 &  1   &   1 & 1.74 & 0.44 & 0.44 & 2    & SER5        & 6   & 1    & 2   & --   & --   & --   & --  \\
K6                  &  2.5 & 0.5  & 0.7 & 1.09 & 0.22 & 0.31 & 0.4  & SER6        & 2.5 & 0.6  & 0.5 & --   & --   & --   & --  \\
K7                  &    5 &   1  & 1.5 & 2.18 & 0.44 & 0.65 & 44   & SER7        & 8   & 1.5  & 1.5 & --   & --   & --   & --  \\
K10                 &   2  & 0.5  & 0.5 & 0.87 & 0.22 & 0.22 & 1.8  &             &     &      &     &      &      &      &     \\
\bottomrule
\end{tabular}
\tablefoot{Columns 2 to 4 and 10 to 12: length $l$ and width $w$ 
of the bow-shock structure, and distance $R$ from the star to the 
midpoint of the bow shock, in angular units. 
The sizes were estimated from WISE band 4 (22 $\mu$m) images, 
and with band 3 (12.1 $\mu$m) images when identification was not 
clear in band 4.
Columns 5 to 7 and 13 to 15: same variables as before, in linear 
units, for those stars of known distances.
Columns 8 and 16: ambient density as derived in \cite{Peri2012} 
\citep{Wilkin1996}.
(*): studied in E-BOSS r1 \citep{Peri2012} only through MSX images.}}
\end{table*}

\subsection{Special cases}

We found several doubtful cases during the BSCs search. All the 
special cases are marked with ** in the Tables. The cases are: 
in Group 3 (Table \ref{grupo3}), HIP 48469 and 111071; in Group 
4 (Table \ref{grupo4}), HIP 3478, 11894, 15114, 30169, 32786, 
34485, 43955, 48589, 62913, 76416, 94385, 99618, and 105268; 
in Group 6 (Table \ref{grupo6}), HIP 102274.

In WISE band 4 (22.2 $\mu$m) many stars show circular emission 
around the stellar object which is usually seen in band 1, 
3.4 $\mu$m. This could indicate that the radial velocity dominates 
over the tangential one. Other cases show different shapes depending 
on the WISE band, which makes us wonder if they were BSC emission 
or something else. In some cases, therefore, the bow shock may still 
be forming and has not yet achieved the comma-shape.

Group 7 (Table \ref{grupo7}) has nine special cases. Two sources 
are from \cite{Kobulnicky2010}: K8 and K9. The stellar bow-shock 
number 8 is mixed with ISM on WISE images and we did not see a clear 
comma-shaped object; K9 has, as many of the special cases in this 
work, circular emission around the star. HD 192281 (\citealt{Arnal2011}) 
images of WISE do not show a clear bow-shape either. The last special 
case is H4. The WISE resolution overcomes the Hubble one, precluding 
a BSC identification. The BSCs related to G3, G4, G6 \citep{Gvaramadze2011b}, 
SER3, and SER4 are doubtful cases on the basis of morphology, but we 
include them on the E-BOSS r2 list because they are candidates.

We do not discard any of the special cases as a probable BSC, but for  
the reasons stated above, we could not identify them. Subsequent studies 
can reveal the sources in the future and help to decide their nature.

\subsection{Statistics}

We carried out statistical studies for Groups 2 to 6 (E-BOSS r1 and r2) 
because they share the same search criterion: we looked for BSCs around
runaway stars. We studied the spatial distribution and spectral types,
and displayed the corresponding plots in Figures 1 and 2. We built 
these plots to look for trends related to stellar spectral types or 
spatial distribution and the presence of BSCs in the star sample.

Figure 1 shows the spatial distribution in galactic coordinates for all 
the sources of Groups 2 to 6, using two different symbols for the objects 
with or without BSCs found by us. From this figure we found no clear spatial 
tendencies for the sources associated with BSCs. 

Figure 2 gives the total number of stars for each subspectral type, 
from O2 to B5, and the proportion of sources with BSCs. We had to 
exclude several cases; for Group 4 (stars B3-B5) we discarded sources 
of B type with no subspectral type, and for Group 3 (Tetzlaff WISE 2) 
we excluded two stars of Op type. It can be seen from the plot that 
the probability of finding a BSC in the surroundings of B-type stars 
($\sim 1.2$ \%) is less than that for O-type stars ($\sim 3.6$ \%).
This has to be taken into account of with caution because the sample 
of B stars in E-BOSS r2 is much larger than in r1.

\subsection{Radio and high-energy emission}

The first evidence of radio synchrotron emission from a stellar bow shock 
was presented by \cite{Benaglia2010} and it was related to the O supergiant 
runaway star BD +43$^{\circ}$ 3654, apparently ejected from the Cygnus OB2 
association (\citealt{Comeron2007}). The bow shock of BD +43$^{\circ}$ 3654 
is the prototypical case, with matching infrared and possibly non-thermal 
radio emission. Synchrotron radiation reveals the presence of accelerated 
electrons plus a magnetic field in the bow-shock region. The accelerated 
electrons give rise to the possible existence of other relativistic particles. 
These particles can produce different non-thermal processes and high-energy 
photons. \cite{Benaglia2010} built a zero-order SED that fits the radio 
emission and predicted the detectability at shorter wavelengths. Different 
observational studies and models have explored this idea on other runaway 
stars:, $\zeta$ Ophiuchi, AE Aurigae, and HD 195592 
(\citealt{Terada2012,delValle2012,Lopez-S2012,delValle2013}).
Non-thermal radiative emission models developed for these cases indicate that 
high-energy photons are mainly produced by the inverse Compton process 
(relativistic electrons interacting with IR dust photons). Moreover, the bow 
shock around HD 195592, is apparently related to the {\it Fermi} source 
2FGL J2030.7+4417 (\citealt{delValle2013}). In addition, runaway stars as 
possible variable gamma-ray sources have been studied by \cite{delValle2014}. 
They found that these stars could form a set of variable gamma-ray sources 
switching on and off in scales of years. We looked for the final BSCs in 
radio databases in order to find more examples that might contribute to a 
better understanding of the radio-gamma connection in this kind of sources.

The NVSS VLA postage service allowed us to search for radio emission at 
1.4 GHz (\citealt{Condon1998}) for almost all the sources. HIP 46928, 
SER 1, 2, and 3, and J1117-6120 were not covered by the NVSS (declination 
below -40 degrees). Many cases show no radio emission in the BSC region, and
many others show confusing emission sources and were discarded as candidates.
A few cases display very interesting features, and are perhaps related to 
the stellar BSCs observed at IR wavelengths. These cases are G2, G3, and 
SER 5. Additionally, we built images from raw data of the Australia 
Telescope Online Archive\footnote{http://atoa.atnf.csiro.au/}
(ATOA, project C492, \citealt{Whiteoak1997}) to examine the RCW 49 zone. 
We found radio emission in the BSCs RCW 49 S1 and S3. Together with three 
cases from E-BOSS r1 that also show radio features (HIP 88652, HIP 38430, 
and HIP 11891), we reunited a total of eight cases with possible radio 
emission coincident with the IR radiation. Images of radio and IR emission 
are shown in Figures 3 and 4. The cases of RCW-49 S1 and S3 were saturated 
on WISE images and are not presented here.

We estimated, using the NVSS 1.4 GHz maps, noise levels and integrated 
fluxes over the IR WISE 4 emission regions for G2, G3, SER5, HIP 11891, 
HIP 24575 (AE Aur), HIP 38430, and HIP 88652. The values of noise and 
integrated fluxes are (in the same order as above): 4 mJy/beam, 1 mJy; 
26 mJy/beam, 2 mJy; 5.5 mJy/beam, 0.5 mJy; 300 mJy/beam, 1 mJy; 1.4 mJy/beam, 
0.25 mJy; 256 mJy/beam, 0.5 mJy; 22 mJy/beam, 0.4 mJy. The areas where we 
integrated the fluxes for NVSS do not always fully match with the IR emission 
exposed by WISE bands, which does not permit us to know whether all the 
radio emission comes from the bow-shock region. For the rest of the regions 
examined the noise level is $\sim$ 0.5 mJy, which tell us that BSCs emission 
is lower or not-existent.

%__________________________________________________________________

\section{Discussion}

Stellar bow shocks are known to be generated by high-mass early-type 
runaway stars, and have been identified not only in the Milky Way 
(e.g., \citealt{Peri2012}, this work) but also in the Small Magellanic 
Cloud \citep{GvaramadzeSMC2011}. Runaway stars emerge from dense 
clusters and are produced by two mechanisms that seem to be functioning 
nowadays (e.g., \citealt{Hoogerwerf2001}): the binary supernova scenario 
(BSS, \citealt{Blaauw1961}) and the dynamical ejection scenario (DES, 
\citealt{Poveda1967}). Runaway stars create the possibility of exploring 
the existence of neutron stars in their surroundings (e.g., 
\citealt{Hoogerwerf2001}). Cases like Vela X-1 \citep{Kaper1997} and 
4U 1907+09 (\citealt{Gvaramadze2008} and this paper) revealed that 
bow shocks can help to seek high-mass X-ray binaries. 

Characterization of the interstellar matter can be done by means 
of stellar bow-shock morphology analysis. Using the stagnation point 
\citep{Wilkin1996,Peri2012}, estimated by examination of the bow-shock 
shape and size, the medium density $n_{\rm ISM}$ can be calculated for 
each region of interest. We calculated $n_{\rm ISM}$ for all the BSCs 
that show emission in WISE band 4 or 3 and show the values in Table 
\ref{medidas}. The values of $n_{\rm ISM}$ are mostly between 0.1 and 
30 cm$^{-3}$, and some sources show high values as 42, 44, 85, and 380 
cm$^{-3}$. The combination of low $R_0$ and high $M_{\odot}$ can be a 
determining factor.

In addition, data of each bow-shock candidate in several infrared 
wavelengths can help to construct spectral energy distributions and 
compare them with dust models to estimate matter temperatures or 
densities (Peri et al., in preparation). 

It is of great importance to compare the density values obtained with 
both methods because each technique has different error sources.  

Many recently discovered high-energy sources (around 30$\%$) have 
no identified counterparts, or more than one in other wavelengths 
(see, e.g., results from {\it Fermi} second catalogue, \citealt{Nolan2012}). 
The stellar bow shock produced by BD +43$^{\circ}$ 3654 was the first 
to be proposed as a possible high-energy source \citep{Benaglia2010}. 
Dedicated papers following this idea have been published (\citealt{Benaglia2010,
Terada2012,delValle2012,Lopez-S2012,delValle2013}), but only the case 
of the BSC related to HD 195592 remains as a possible high-energy 
({\it Fermi}) source. Recently, a thorough study of {\it Fermi} data 
has been carried out by \cite{Schulz2014} for all the BSCs of E-BOSS 
r1 \citep{Peri2012}. The authors found no evidence of $\gamma$-ray 
emission in any case, and they attribute this to the possible inefficient 
particle acceleration or to the fact that the photon density coming 
from the dust is lower than assumed. \cite{Terada2012}, who have found 
no diffuse X-ray emission for BD +43$^{\circ}$ 3654 but only an upper 
limit, have pointed out that it can be consequence of the low magnetic 
field turbulence.

Future radio observations from instruments like the Very Large 
Array\footnote{https://science.nrao.edu/facilities/vla} (VLA) or Giant 
Metrewave Radio Telescope\footnote{http://gmrt.ncra.tifr.res.in/} 
(GMRT) of the eight BSCs where we encountered NVSS 1.4 GHz emission 
features can help to obtain synchrotron emission parameters such as 
spectral indexes, and can contribute to finding better constraints 
of the non-thermal radiative models. In addition, deeper observations 
performed with instruments like {\it Fermi} or future ground-based 
arrays like CTA can give new information about the high-energy 
emission of stellar bow shocks and add information in similar 
studies.

%____________________________________________________________________________

\section{Summary and conclusions}

We discovered 16 new stellar bow-shock candidates and gathered 29 
additional from recent literature. The 45 objects, together with 
those from \cite{Peri2012}, form the Extensive stellar BOw Shock 
Survey second release. Whenever the identification was possible 
we present the stellar parameters of the runaway stars that probably 
generated them. We list the parameters of all bow-shock candidates. 
The complete E-BOSS list contains 73 objects.

Most of the bow-shock search consisted of investigating the surroundings 
of runaway stars (Groups 2 to 6). From 503 of such objects we found 
27 BSCs, which represents a 5.4$\%$ rate of success. This proportion is 
lower than that obtained in E-BOSS r1, in which, of a total of 164 
runaway stars, we encountered 17 BSCs, a success rate of 10$\%$ 
(only Group 2). This difference can be explained by the higher number 
of B-type stars sampled in this paper (Groups 2 to 6, 84 O-stars 
and 486 B-stars; this is in agreement with the expected number of 
stars of spectral types O and B for the general case). We conclude 
that the proportion of BSCs among O-type stars ($\sim$ 3.6 \%) is 
larger than for B-type stars ($\sim$ 1.2 \%). This can be associated 
with the high values of several parameters for O-type stars, such as 
mass-loss rates, intense UV photon fields, temperature, and wind 
terminal velocity. Moreover, physical conditions of the ISM, as well 
as the relative velocities of the star and medium can affect the 
shape or even allow or preclude the formation of the bow shock.

The sizes of the BSCs are similar to those observed on E-BOSS r1. 
The shapes are variable: typical bow shape (e.g., K5, BD +43$^{\circ}$ 
3654, or HIP 101186), layered shells (SER 1, for example), asymmetric,
or cases that might be showing evidence of ISM inhomogeneities.

Further exploration of IR databases and new observations of instruments 
like Gaia\footnote{http://sci.esa.int/gaia}\ will contribute to enlarging 
the sample of stellar bow shock and runaway stars.

%________________________________________________________________________________________________

\begin{table*}
\caption{Total list of BSC of E-BOSS r1 and r2.}
\label{tabla2versiones}
{\centering
\begin{tabular}{| l l c c | l l c c | l l c c |}
\toprule
ID   & Name      & Rel.& Group  & ID  & Name        & Rel.& Group & ID & Name         & Rel.& Group \\
\midrule 
EB01 & HIP 2036  & r1 & 2       & EB26 & HIP 101186 & r1 & 1	& EB51 & HIP 86768    & r2 & 6 \\
EB02 & HIP 2599  & r1 & 1,2	& EB27 & BD+433654  & r1 & 1	& EB52 & Star 1       & r2 & 7 \\
EB03 & HIP 11891 & r1 & 2	& EB28 & HIP 114990 & r1 & 1 	& EB53 & M17-S1       & r2 & 7 \\
EB04 & HIP 16518 & r1 & 2	& EB29 & SER6       & r2 & 7	& EB54 & M17-S2       & r2 & 7 \\
EB05 & HIP 17358 & r1 & 1	& EB30 & SER5       & r2 & 7	& EB55 & M17-S3       & r2 & 7 \\
EB06 & HIP 22783 & r1 & 1	& EB31 & HIP 57862  & r2 & 5 	& EB56 & BD -14 5040  & r2 & 7 \\
EB07 & HIP 24575 & r1 & 2	& EB32 & SER1       & r2 & 7	& EB57 & 4U 1907+09   & r2 & 7 \\
EB08 & HIP 25923 & r1 & 2	& EB33 & HIP 44368  & r2 & 3,7	& EB58 & HIP 98418    & r2 & 3 \\
EB09 & HIP 26397 & r1 & 2	& EB34 & HIP 46928  & r2 & 4 	& EB59 & TYC 3159-6-1 & r2 & 7 \\
EB10 & HIP 28881 & r1 & 1	& EB35 & HIP 47868  & r2 & 3 	& EB60 & K8           & r2 & 7 \\
EB11 & HIP 29276 & r1 & 1,2	& EB36 & SER2       & r2 & 7	& EB61 & K9           & r2 & 7 \\
EB12 & HIP 31766 & r1 & 2	& EB37 & RCW 49-S1  & r2 & 7	& EB62 & K3           & r2 & 7 \\
EB13 & HIP 32067 & r1 & 1,2	& EB38 & RCW 49-S2  & r2 & 7	& EB63 & K4           & r2 & 7 \\
EB14 & HIP 34536 & r1 & 1,2	& EB39 & RCW 49-S3  & r2 & 7	& EB64 & K10          & r2 & 7 \\
EB15 & HIP 38430 & r1 & 1	& EB40 & SER3       & r2 & 7	& EB65 & K1           & r2 & 7 \\
EB16 & HIP 62322 & r1 & 2	& EB41 & J1117-6120 & r2 & 7	& EB66 & K2           & r2 & 7 \\
EB17 & HIP 72510 & r1 & 1,2	& EB42 & SER7       & r2 & 7	& EB67 & K5           & r2 & 7 \\
EB18 & HIP 75095 & r1 & 1,2	& EB43 & G4         & r2 & 7	& EB68 & K7           & r2 & 7 \\
EB19 & HIP 77391 & r1 & 1	& EB44 & G2         & r2 & 7	& EB69 & K6           & r2 & 7 \\
EB20 & HIP 78401 & r1 & 1	& EB45 & G5         & r2 & 7	& EB70 & HIP 104579   & r2 & 3 \\
EB21 & HIP 81377 & r1 & 1,2	& EB46 & G6         & r2 & 7	& EB71 & HIP 105186   & r2 & 3 \\
EB22 & HIP 82171 & r1 & 2	& EB47 & G8         & r2 & 7	& EB72 & HIP 107789   & r2 & 4 \\
EB23 & HIP 88652 & r1 & 2	& EB48 & G1         & r2 & 7	& EB73 & SER4         & r2 & 7 \\ 
EB24 & HIP 92865 & r1 & 1	& EB49 & G7         & r2 & 7	&      &              &    &   \\
EB25 & HIP 97796 & r1 & 1 	& EB50 & G3         & r2 & 7	&      &              &    &   \\    
\bottomrule
\end{tabular}
\tablefoot{Bow-shock candidates of E-BOSS release 1 and 2 reunited. 
We give a specific name to the catalogue that goes from 01 to 73 with EB from E-BOSS as a prefix.
There are 28 objects from r1 and 45 from r2. The Group for each source is also given.}}
\end{table*}

%______________________________________________________________________________________________

\begin{acknowledgements}
  C.S. Peri and P.B. are supported by the ANPCyT PICT-2012/00878. P.B. 
also acknowledges support from CONICET PIP 0078 and UNLP G11/115
projects. This publication makes uses of the 
NASA/IPAC Infrared Science Archive, which is operated by the Jet 
Propulsion Laboratory, California Institute of Technology, under 
contract with the National Aeronautics and Space Administration.
We thank the SIMBAD database, operated at CDS, Strasbourg, France, and data 
products from the Wide-field Infrared Survey Explorer, which is a 
joint project of the University of California, Los Angeles, and 
the Jet Propulsion Laboratory/California Institute of Technology, 
funded by the National Aeronautics and Space Administration. 
We also thank the HubbleSite, which is produced by the Space 
Telescope Science Institute (STScI). STScI is operated by the 
Association of Universities for Research in Astronomy, Inc. (AURA) 
for NASA, under contract with the Goddard Space Flight Center, 
Greenbelt, MD. The Hubble Space Telescope is a project of 
international cooperation between NASA and the European Space 
Agency (ESA). We thank the Hubble Legacy Archive, which is a 
collaboration between the Space Telescope Science Institute (STScI/NASA), 
the Space Telescope European Coordinating Facility (ST-ECF/ESA) and 
the Canadian Astronomy Data Centre (CADC/NRC/CSA).
Finally, we appreciate the suggestions of A. Noriega-Crespo, and
insightful discussions with M. V. del Valle and L. J. Pellizza.
We thank the anonymous A\&A referee for comments and suggestions that
improved the article.

\end{acknowledgements}

%-------------------------------------------------------------------

\bibliography{eboss2.v1.bib}{}

\begin{thebibliography}{56}
\expandafter\ifx\csname natexlab\endcsname\relax\def\natexlab#1{#1}\fi

\bibitem[{{Ankay} {et~al.}(2001){Ankay}, {Kaper}, {de Bruijne}, {Dewi},
  {Hoogerwerf}, \& {Savonije}}]{Ankay2001}
{Ankay}, A., {Kaper}, L., {de Bruijne}, J.~H.~J., {et~al.} 2001, \aap, 370, 170

\bibitem[{{Arnal} {et~al.}(2011){Arnal}, {Cichowolski}, {Pineault}, {Testori},
  \& {Cappa}}]{Arnal2011}
{Arnal}, E.~M., {Cichowolski}, S., {Pineault}, S., {Testori}, J.~C., \&
  {Cappa}, C.~E. 2011, \aap, 532, A9

\bibitem[{{Benaglia}(2012)}]{Benaglia2012}
{Benaglia}, P. 2012, Boletin de la Asociacion Argentina de Astronomia La Plata
  Argentina, 55, 43

\bibitem[{{Benaglia} {et~al.}(2010){Benaglia}, {Romero}, {Mart{\'{\i}}},
  {Peri}, \& {Araudo}}]{Benaglia2010}
{Benaglia}, P., {Romero}, G.~E., {Mart{\'{\i}}}, J., {Peri}, C.~S., \&
  {Araudo}, A.~T. 2010, \aap, 517, L10

\bibitem[{{Benjamin} {et~al.}(2003){Benjamin}, {Churchwell}, {Babler}, {Bania},
  {Clemens}, {Cohen}, {Dickey}, {Indebetouw}, {Jackson}, {Kobulnicky},
  {Lazarian}, {Marston}, {Mathis}, {Meade}, {Seager}, {Stolovy}, {Watson},
  {Whitney}, {Wolff}, \& {Wolfire}}]{Benjamin2003}
{Benjamin}, R.~A., {Churchwell}, E., {Babler}, B.~L., {et~al.} 2003, \pasp,
  115, 953

\bibitem[{{Blaauw}(1961)}]{Blaauw1961}
{Blaauw}, A. 1961, \bain, 15, 265

\bibitem[{{Brown} \& {Bomans}(2005)}]{Brown2005}
{Brown}, D. \& {Bomans}, D.~J. 2005, \aap, 439, 183

\bibitem[{{Comer{\'o}n} \& {Pasquali}(2007)}]{Comeron2007}
{Comer{\'o}n}, F. \& {Pasquali}, A. 2007, \aap, 467, L23

\bibitem[{{Condon} {et~al.}(1998){Condon}, {Cotton}, {Greisen}, {Yin},
  {Perley}, {Taylor}, \& {Broderick}}]{Condon1998}
{Condon}, J.~J., {Cotton}, W.~D., {Greisen}, E.~W., {et~al.} 1998, \aj, 115,
  1693

\bibitem[{{Cruz-Gonz{\'a}lez} {et~al.}(1974){Cruz-Gonz{\'a}lez},
  {Recillas-Cruz}, {Costero}, {Peimbert}, \&
  {Torres-Peimbert}}]{CruzGonzalez1974}
{Cruz-Gonz{\'a}lez}, C., {Recillas-Cruz}, E., {Costero}, R., {Peimbert}, M., \&
  {Torres-Peimbert}, S. 1974, \rmxaa, 1, 211

\bibitem[{{del Valle} \& {Romero}(2012)}]{delValle2012}
{del Valle}, M.~V. \& {Romero}, G.~E. 2012, \aap, 543, A56

\bibitem[{{del Valle} \& {Romero}(2014)}]{delValle2014}
{del Valle}, M.~V. \& {Romero}, G.~E. 2014, \aap, 563, A96

\bibitem[{{del Valle} {et~al.}(2013){del Valle}, {Romero}, \& {De
  Becker}}]{delValle2013}
{del Valle}, M.~V., {Romero}, G.~E., \& {De Becker}, M. 2013, \aap, 550, A112

\bibitem[{{Egan} {et~al.}(2003){Egan}, {Price}, \& {Kraemer}}]{Egan2003}
{Egan}, M.~P., {Price}, S.~D., \& {Kraemer}, K.~E. 2003, in Bulletin of the
  American Astronomical Society, Vol.~35, American Astronomical Society Meeting
  Abstracts, 1301

\bibitem[{{Fujii} \& {Portegies Zwart}(2011)}]{Fujii2011}
{Fujii}, M.~S. \& {Portegies Zwart}, S. 2011, Science, 334, 1380

\bibitem[{{Gies}(1987)}]{Gies1987}
{Gies}, D.~R. 1987, \apjs, 64, 545

\bibitem[{{Gies} \& {Bolton}(1986)}]{Gies1986}
{Gies}, D.~R. \& {Bolton}, C.~T. 1986, \apjs, 61, 419

\bibitem[{{Gull} \& {Sofia}(1979)}]{Gull1979}
{Gull}, T.~R. \& {Sofia}, S. 1979, \apj, 230, 782

\bibitem[{{Gvaramadze} \& {Bomans}(2008)}]{Gvaramadze2008}
{Gvaramadze}, V.~V. \& {Bomans}, D.~J. 2008, \aap, 490, 1071

\bibitem[{{Gvaramadze} {et~al.}(2013){Gvaramadze}, {Kniazev}, {Chen{\'e}}, \&
  {Schnurr}}]{Gvaramadze2013}
{Gvaramadze}, V.~V., {Kniazev}, A.~Y., {Chen{\'e}}, A.-N., \& {Schnurr}, O.
  2013, \mnras, 430, L20

\bibitem[{{Gvaramadze} {et~al.}(2011{\natexlab{a}}){Gvaramadze}, {Kniazev},
  {Kroupa}, \& {Oh}}]{Gvaramadze2011b}
{Gvaramadze}, V.~V., {Kniazev}, A.~Y., {Kroupa}, P., \& {Oh}, S.
  2011{\natexlab{a}}, \aap, 535, A29

\bibitem[{{Gvaramadze} {et~al.}(2014){Gvaramadze}, {Miroshnichenko}, {Castro},
  {Langer}, \& {Zharikov}}]{Gvaramadze2014}
{Gvaramadze}, V.~V., {Miroshnichenko}, A.~S., {Castro}, N., {Langer}, N., \&
  {Zharikov}, S.~V. 2014, \mnras, 437, 2761

\bibitem[{{Gvaramadze} {et~al.}(2011{\natexlab{b}}){Gvaramadze},
  {Pflamm-Altenburg}, \& {Kroupa}}]{GvaramadzeSMC2011}
{Gvaramadze}, V.~V., {Pflamm-Altenburg}, J., \& {Kroupa}, P.
  2011{\natexlab{b}}, \aap, 525, A17

\bibitem[{{Gvaramadze} {et~al.}(2011{\natexlab{c}}){Gvaramadze}, {R{\"o}ser},
  {Scholz}, \& {Schilbach}}]{Gvaramadze2011a}
{Gvaramadze}, V.~V., {R{\"o}ser}, S., {Scholz}, R.-D., \& {Schilbach}, E.
  2011{\natexlab{c}}, \aap, 529, A14

\bibitem[{{Hoogerwerf} {et~al.}(2001){Hoogerwerf}, {de Bruijne}, \& {de
  Zeeuw}}]{Hoogerwerf2001}
{Hoogerwerf}, R., {de Bruijne}, J.~H.~J., \& {de Zeeuw}, P.~T. 2001, \aap, 365,
  49

\bibitem[{{Kaper} {et~al.}(1997){Kaper}, {van Loon}, {Augusteijn},
  {Goudfrooij}, {Patat}, {Waters}, \& {Zijlstra}}]{Kaper1997}
{Kaper}, L., {van Loon}, J.~T., {Augusteijn}, T., {et~al.} 1997, \apjl, 475,
  L37

\bibitem[{{Kharchenko} {et~al.}(2007){Kharchenko}, {Scholz}, {Piskunov},
  {R{\"o}ser}, \& {Schilbach}}]{Kharchenko2007}
{Kharchenko}, N.~V., {Scholz}, R.-D., {Piskunov}, A.~E., {R{\"o}ser}, S., \&
  {Schilbach}, E. 2007, Astronomische Nachrichten, 328, 889

\bibitem[{{Kobulnicky} {et~al.}(2010){Kobulnicky}, {Gilbert}, \&
  {Kiminki}}]{Kobulnicky2010}
{Kobulnicky}, H.~A., {Gilbert}, I.~J., \& {Kiminki}, D.~C. 2010, \apj, 710, 549

\bibitem[{{Liu} {et~al.}(2011){Liu}, {Padgett}, {Leisawitz}, {Fajardo-Acosta},
  \& {Koenig}}]{Liu2011}
{Liu}, W.~M., {Padgett}, D.~L., {Leisawitz}, D., {Fajardo-Acosta}, S., \&
  {Koenig}, X.~P. 2011, \apjl, 733, L2

\bibitem[{{L{\'o}pez-Santiago} {et~al.}(2012){L{\'o}pez-Santiago}, {Miceli},
  {del Valle}, {Romero}, {Bonito}, {Albacete-Colombo}, {Pereira}, {de Castro},
  \& {Damiani}}]{Lopez-S2012}
{L{\'o}pez-Santiago}, J., {Miceli}, M., {del Valle}, M.~V., {et~al.} 2012,
  \apjl, 757, L6

\bibitem[{{Magnier} {et~al.}(1999){Magnier}, {Volp}, {Laan}, {van den Ancker},
  \& {Waters}}]{Magnier1999}
{Magnier}, E.~A., {Volp}, A.~W., {Laan}, K., {van den Ancker}, M.~E., \&
  {Waters}, L.~B.~F.~M. 1999, \aap, 352, 228

\bibitem[{{Ma{\'{\i}}z-Apell{\'a}niz}
  {et~al.}(2004){Ma{\'{\i}}z-Apell{\'a}niz}, {Walborn}, {Galu{\'e}}, \&
  {Wei}}]{Maiz2004}
{Ma{\'{\i}}z-Apell{\'a}niz}, J., {Walborn}, N.~R., {Galu{\'e}}, H.~{\'A}., \&
  {Wei}, L.~H. 2004, \apjs, 151, 103

\bibitem[{{Mason} {et~al.}(1998){Mason}, {Gies}, {Hartkopf}, {Bagnuolo}, {ten
  Brummelaar}, \& {McAlister}}]{Mason1998}
{Mason}, B.~D., {Gies}, D.~R., {Hartkopf}, W.~I., {et~al.} 1998, \aj, 115, 821

\bibitem[{{Megier} {et~al.}(2009){Megier}, {Strobel}, {Galazutdinov}, \&
  {Kre{\l}owski}}]{Megier2009}
{Megier}, A., {Strobel}, A., {Galazutdinov}, G.~A., \& {Kre{\l}owski}, J. 2009,
  \aap, 507, 833

\bibitem[{{Moffat} {et~al.}(1998){Moffat}, {Marchenko}, {Seggewiss}, {van der
  Hucht}, {Schrijver}, {Stenholm}, {Lundstrom}, {Setia Gunawan}, {Sutantyo},
  {van den Heuvel}, {de Cuyper}, \& {Gomez}}]{Moffat1998}
{Moffat}, A.~F.~J., {Marchenko}, S.~V., {Seggewiss}, W., {et~al.} 1998, \aap,
  331, 949

\bibitem[{{Neugebauer} {et~al.}(1984){Neugebauer}, {Habing}, {van Duinen},
  {Aumann}, {Baud}, {Beichman}, {Beintema}, {Boggess}, {Clegg}, {de Jong},
  {Emerson}, {Gautier}, {Gillett}, {Harris}, {Hauser}, {Houck}, {Jennings},
  {Low}, {Marsden}, {Miley}, {Olnon}, {Pottasch}, {Raimond}, {Rowan-Robinson},
  {Soifer}, {Walker}, {Wesselius}, \& {Young}}]{Neug1984}
{Neugebauer}, G., {Habing}, H.~J., {van Duinen}, R., {et~al.} 1984, \apjl, 278,
  L1

\bibitem[{{Nolan} {et~al.}(2012){Nolan}, {Abdo}, {Ackermann}, {Ajello},
  {Allafort}, {Antolini}, {Atwood}, {Axelsson}, {Baldini}, {Ballet}, \&
  et~al.}]{Nolan2012}
{Nolan}, P.~L., {Abdo}, A.~A., {Ackermann}, M., {et~al.} 2012, \apjs, 199, 31

\bibitem[{{Noriega-Crespo} {et~al.}(1997){Noriega-Crespo}, {van Buren}, \&
  {Dgani}}]{Noriega-Crespo1997}
{Noriega-Crespo}, A., {van Buren}, D., \& {Dgani}, R. 1997, \aj, 113, 780

\bibitem[{{Peri} {et~al.}(2012){Peri}, {Benaglia}, {Brookes}, {Stevens}, \&
  {Isequilla}}]{Peri2012}
{Peri}, C.~S., {Benaglia}, P., {Brookes}, D.~P., {Stevens}, I.~R., \&
  {Isequilla}, N.~L. 2012, \aap, 538, A108

\bibitem[{{Poveda} {et~al.}(1967){Poveda}, {Ruiz}, \& {Allen}}]{Poveda1967}
{Poveda}, A., {Ruiz}, J., \& {Allen}, C. 1967, Boletin de los Observatorios
  Tonantzintla y Tacubaya, 4, 86

\bibitem[{{Povich} {et~al.}(2008){Povich}, {Benjamin}, {Whitney}, {Babler},
  {Indebetouw}, {Meade}, \& {Churchwell}}]{Povich2008}
{Povich}, M.~S., {Benjamin}, R.~A., {Whitney}, B.~A., {et~al.} 2008, \apj, 689,
  242

\bibitem[{{Prinja} {et~al.}(1990){Prinja}, {Barlow}, \& {Howarth}}]{Prinja1990}
{Prinja}, R.~K., {Barlow}, M.~J., \& {Howarth}, I.~D. 1990, \apj, 361, 607

\bibitem[{{Sahai} {et~al.}(2007){Sahai}, {Morris}, {S{\'a}nchez Contreras}, \&
  {Claussen}}]{Sahai2007}
{Sahai}, R., {Morris}, M., {S{\'a}nchez Contreras}, C., \& {Claussen}, M. 2007,
  \aj, 134, 2200

\bibitem[{{Sahai} {et~al.}(2012){Sahai}, {Morris}, \& {Claussen}}]{Sahai2012}
{Sahai}, R., {Morris}, M.~R., \& {Claussen}, M.~J. 2012, \apj, 751, 69

\bibitem[{{Schulz} {et~al.}(2014){Schulz}, {Ackermann}, {Buehler}, {Mayer}, \&
  {Klepser}}]{Schulz2014}
{Schulz}, A., {Ackermann}, M., {Buehler}, R., {Mayer}, M., \& {Klepser}, S.
  2014, \aap, 565, A95

\bibitem[{{Stone}(1991)}]{Stone1991}
{Stone}, R.~C. 1991, \aj, 102, 333

\bibitem[{{Taylor} {et~al.}(2003){Taylor}, {Gibson}, {Peracaula}, {Martin},
  {Landecker}, {Brunt}, {Dewdney}, {Dougherty}, {Gray}, {Higgs}, {Kerton},
  {Knee}, {Kothes}, {Purton}, {Uyaniker}, {Wallace}, {Willis}, \&
  {Durand}}]{Taylor2003}
{Taylor}, A.~R., {Gibson}, S.~J., {Peracaula}, M., {et~al.} 2003, \aj, 125,
  3145

\bibitem[{{Terada} {et~al.}(2012){Terada}, {Tashiro}, {Bamba}, {Yamazaki},
  {Kouzu}, {Koyama}, \& {Seta}}]{Terada2012}
{Terada}, Y., {Tashiro}, M.~S., {Bamba}, A., {et~al.} 2012, \pasj, 64, 138

\bibitem[{{Tetzlaff} {et~al.}(2011){Tetzlaff}, {Neuh{\"a}user}, \&
  {Hohle}}]{Tetzlaff2011}
{Tetzlaff}, N., {Neuh{\"a}user}, R., \& {Hohle}, M.~M. 2011, \mnras, 410, 190

\bibitem[{{van Buren} \& {McCray}(1988)}]{vanBuren1988}
{van Buren}, D. \& {McCray}, R. 1988, \apjl, 329, L93

\bibitem[{{van der Hucht}(2001)}]{vanderHucht2001}
{van der Hucht}, K.~A. 2001, \nar, 45, 135

\bibitem[{{van Leeuwen}(2007)}]{vanLeeuwen2007}
{van Leeuwen}, F. 2007, \aap, 474, 653

\bibitem[{{Vink} {et~al.}(2001){Vink}, {de Koter}, \& {Lamers}}]{Vink2001}
{Vink}, J.~S., {de Koter}, A., \& {Lamers}, H.~J.~G.~L.~M. 2001, \aap, 369, 574

\bibitem[{{Whiteoak} \& {Uchida}(1997)}]{Whiteoak1997}
{Whiteoak}, J.~B.~Z. \& {Uchida}, K.~I. 1997, \aap, 317, 563

\bibitem[{{Wilkin}(1996)}]{Wilkin1996}
{Wilkin}, F.~P. 1996, \apjl, 459, L31

\bibitem[{{Wright} {et~al.}(2010){Wright}, {Eisenhardt}, {Mainzer}, {Ressler},
  {Cutri}, {Jarrett}, {Kirkpatrick}, {Padgett}, {McMillan}, {Skrutskie},
  {Stanford}, {Cohen}, {Walker}, {Mather}, {Leisawitz}, {Gautier}, {McLean},
  {Benford}, {Lonsdale}, {Blain}, {Mendez}, {Irace}, {Duval}, {Liu}, {Royer},
  {Heinrichsen}, {Howard}, {Shannon}, {Kendall}, {Walsh}, {Larsen}, {Cardon},
  {Schick}, {Schwalm}, {Abid}, {Fabinsky}, {Naes}, \& {Tsai}}]{Wright2010}
{Wright}, E.~L., {Eisenhardt}, P.~R.~M., {Mainzer}, A.~K., {et~al.} 2010, \aj,
  140, 1868

\end{thebibliography}
\bibliographystyle{aa.bst}

%-------------------------------------------------------------------

\begin{figure*}[t]
\begin{minipage}{0.49\textwidth}
{\centering
\includegraphics[width=0.85\textwidth]{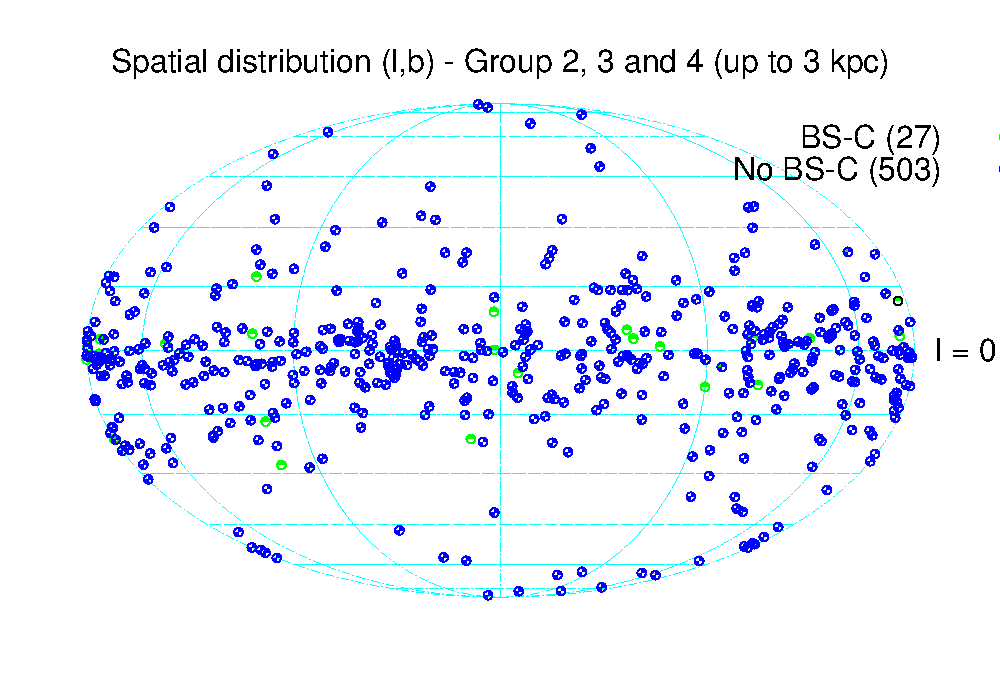}
\caption{Distribution on the (l,b) plane of Groups 2, 3, 4, 5 and 6. Runaway stars were 
  taken from \cite{Tetzlaff2011} and \cite{Hoogerwerf2001}.}}
\label{dist-esp}
\end{minipage} 
\hfill
\begin{minipage}{0.49\textwidth}
{\centering
\includegraphics[width=0.85\textwidth]{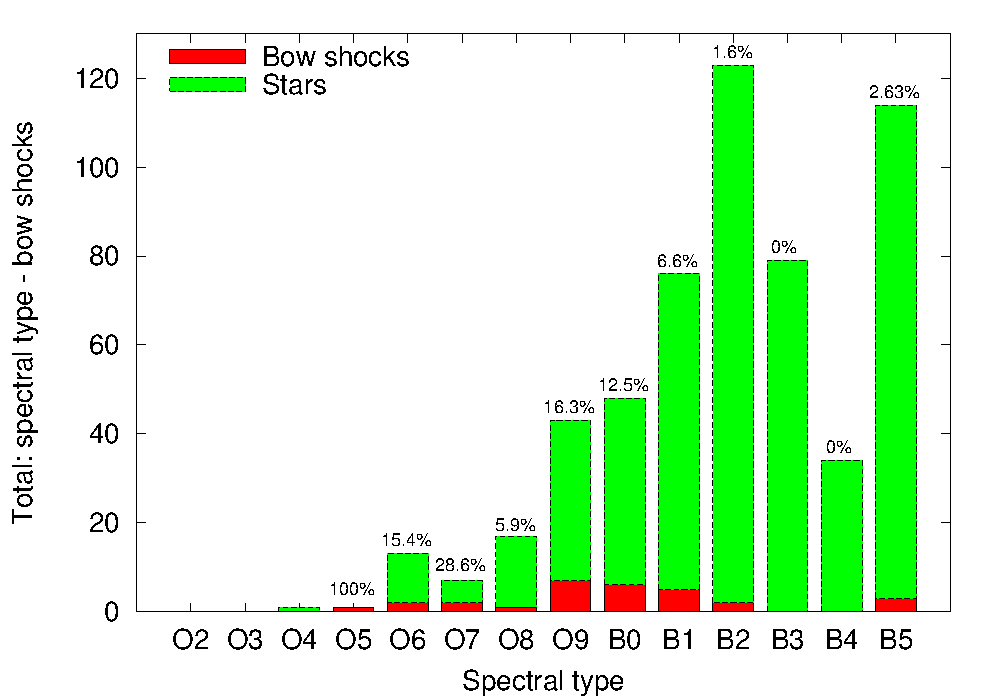}
\caption{Spectral type histogram for Groups 2 to 6, summarized. Each range of spectral types has the
  total number of stars in green, the amount of BSCs in red, and the proportion of BSCs according to the 
  total number of sources in each range. We did not include stars of spectral types with no specific subtype 
  (two stars of spectral type Op, and 12 stars of spectral type B).}}
\label{tipos-esp}
\end{minipage}
\end{figure*} 

%-------------------------------------------------------------------

\begin{figure*}[t]

\begin{minipage}{\textwidth}
\centering
\includegraphics[width=0.42\textwidth]{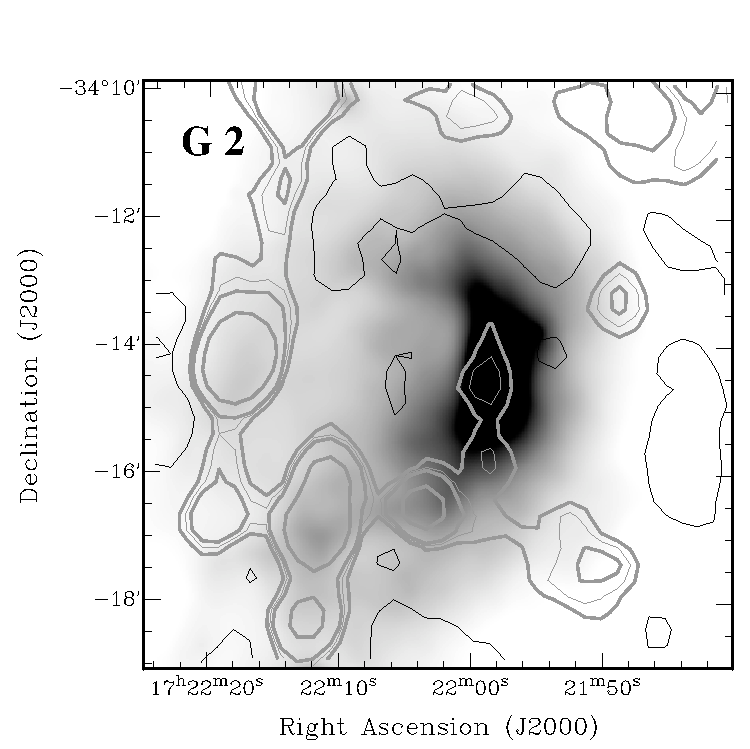}
\hfill
\includegraphics[width=0.44\textwidth]{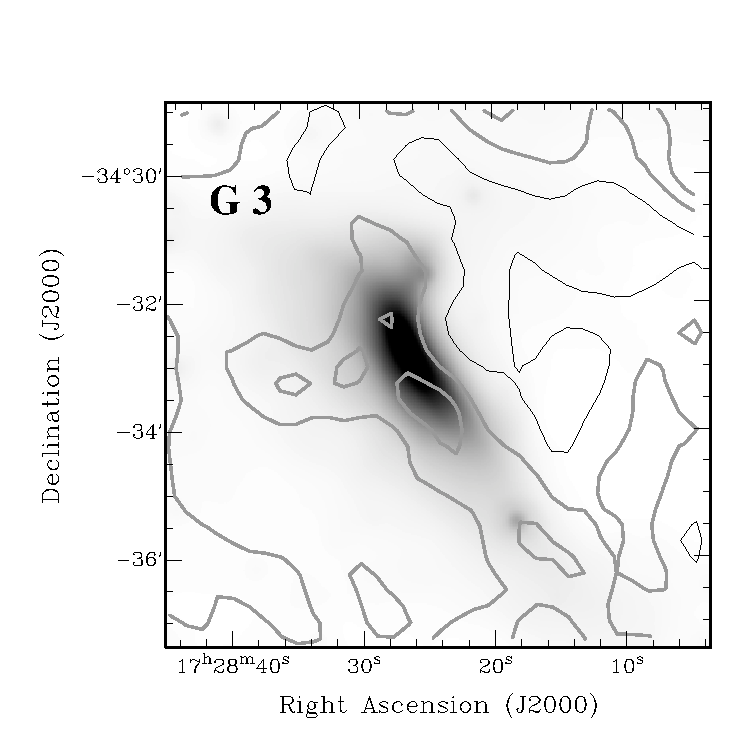}
\end{minipage}

\begin{minipage}{\textwidth}
\centering
\includegraphics[width=0.38\textwidth]{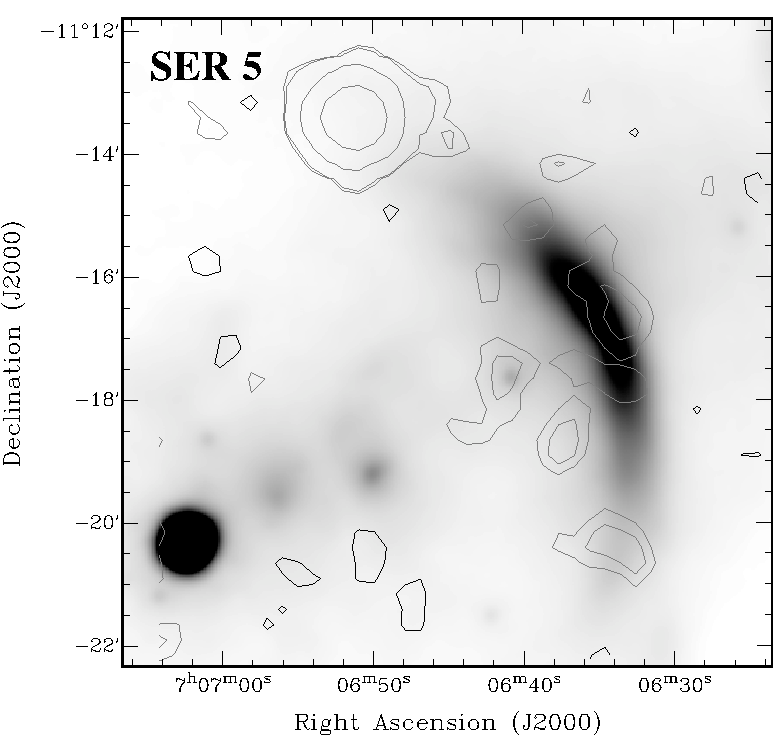}
\hfill
\includegraphics[width=0.38\textwidth]{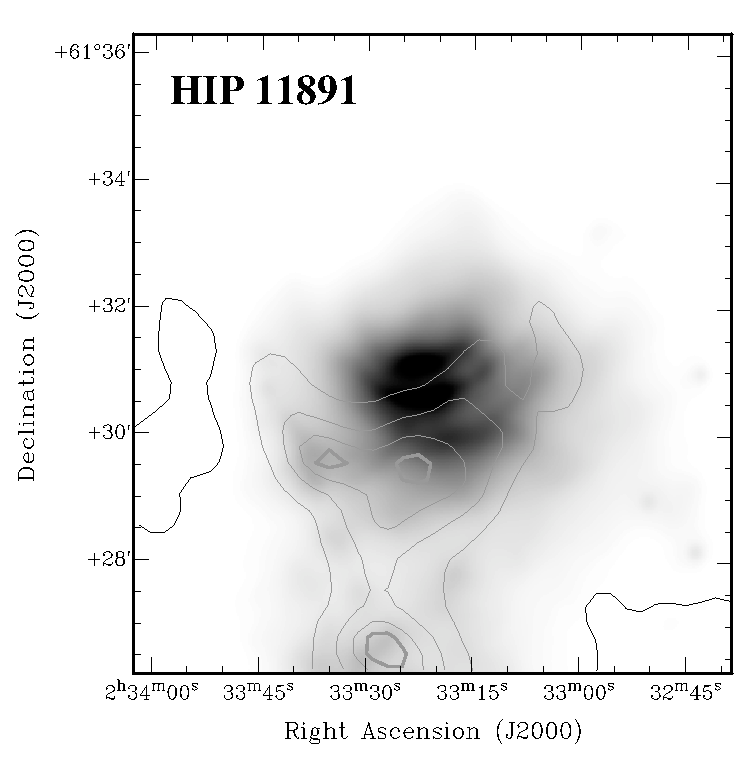}
\end{minipage}

\begin{minipage}{\textwidth}
\centering
\includegraphics[width=0.38\textwidth]{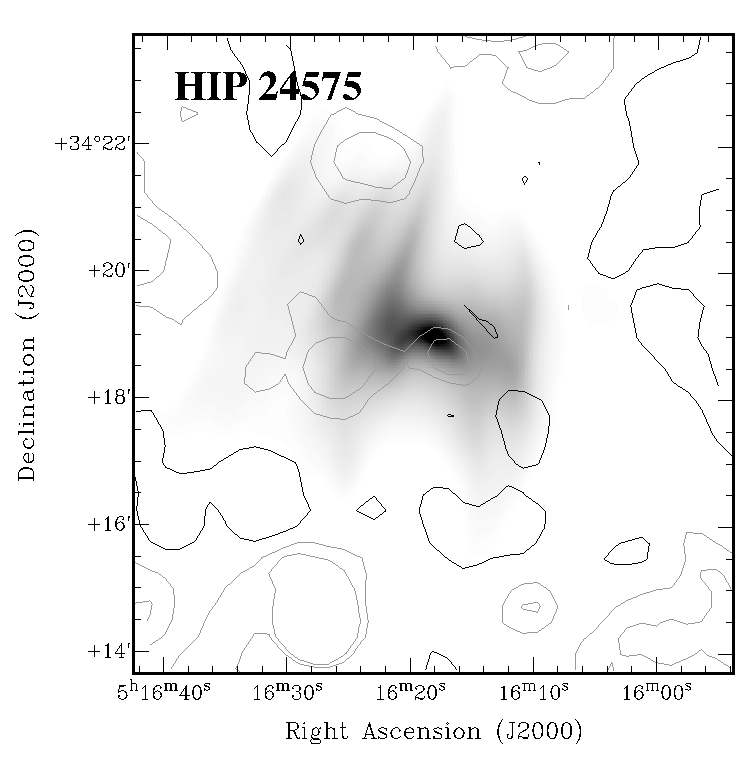}
\hfill
\includegraphics[width=0.38\textwidth]{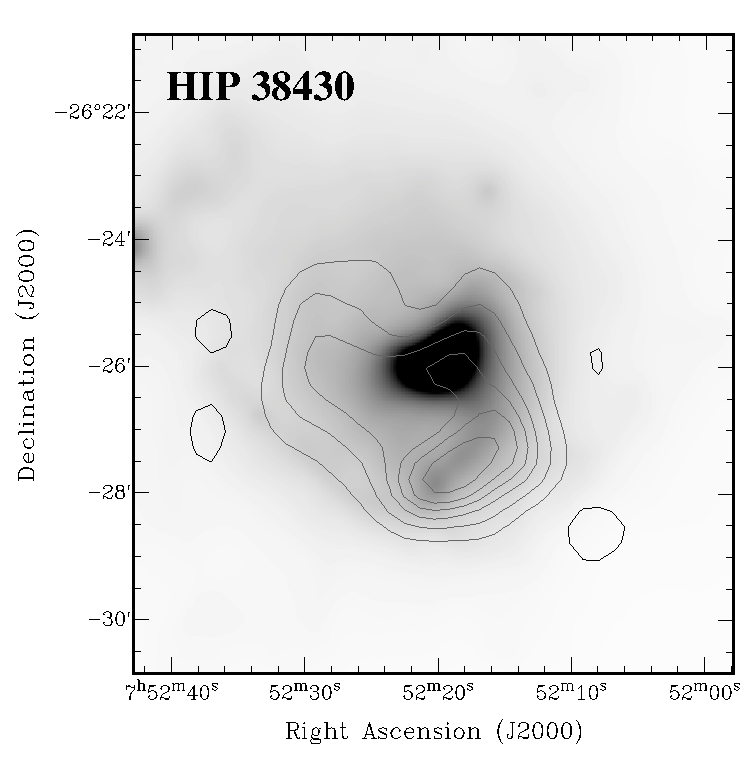}
\end{minipage}

\caption{BSCs that show radio emission.
Gray scale colors: IR WISE emission (band 4, 22.2 $\mu$m). 
Contours: NVSS 1.4 GHz emission. Positive values in gray, negative in black.
Contour levels on each image: 
{\it Top left}, -1, 1, 1.5, 2, 5 mJy/b. 
{\it Top right}, -5, 1, 5 mJy/b. 
{\it Middle left}, -1, 1, 1.5, 20, 150 mJy/b. 
{\it Middle right}, -4, 4, 16, 30, 40 mJy/b. 
{\it Bottom left}, -1, 1, 2, 3 mJy/b. 
{\it Bottom right}, -10, 10, 30, 60, 90, 120, 160 mJy/b.}

\end{figure*}

\begin{figure*}[t]
\centering
\includegraphics[width=0.3\textwidth]{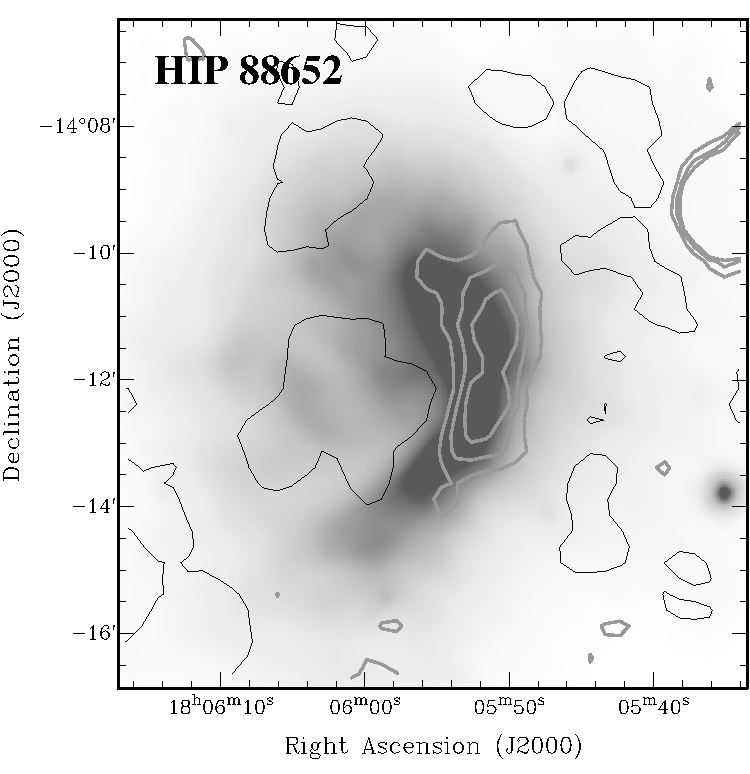}
\caption{Bow-shock candidate related to HIP 88652 (E-BOSS r1).
  Gray scale colors: IR WISE emission (band 4, 22.2 $\mu$m). 
  Contours: NVSS 1.4 GHz emission. Contour levels: -1, 1, 2, 3 mJy/b.
  Positive values in gray, negative in black.}
\end{figure*}

%-------------------------------------------------------------------

\begin{figure*}[t]

\begin{minipage}{\textwidth}
\centering
\includegraphics[width=0.455\textwidth]{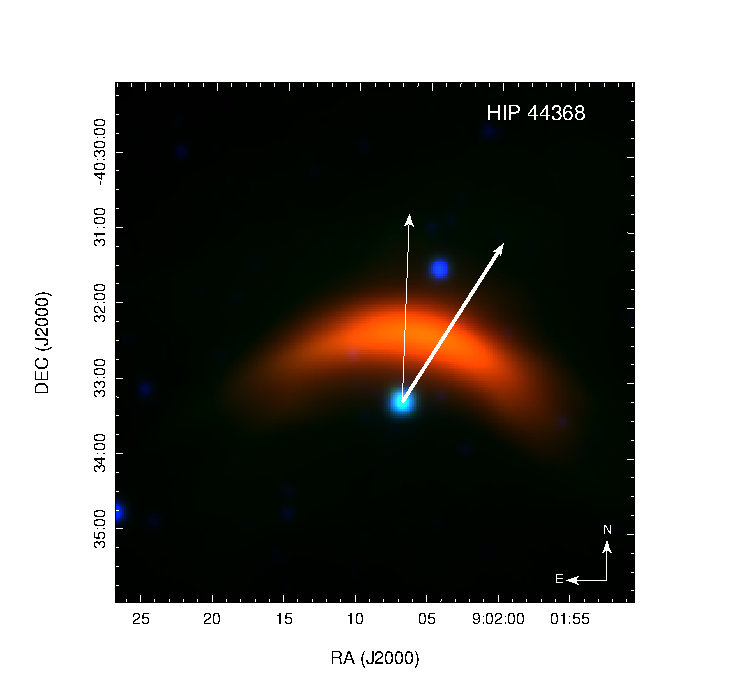}
\hfill
\includegraphics[width=0.455\textwidth]{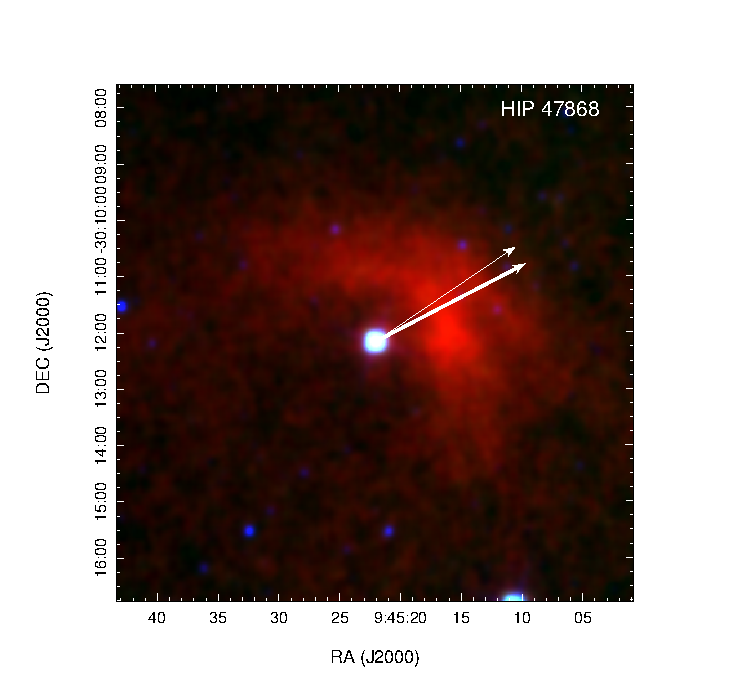}
\end{minipage} 

\begin{minipage}{\textwidth}
\centering
\includegraphics[width=0.455\textwidth]{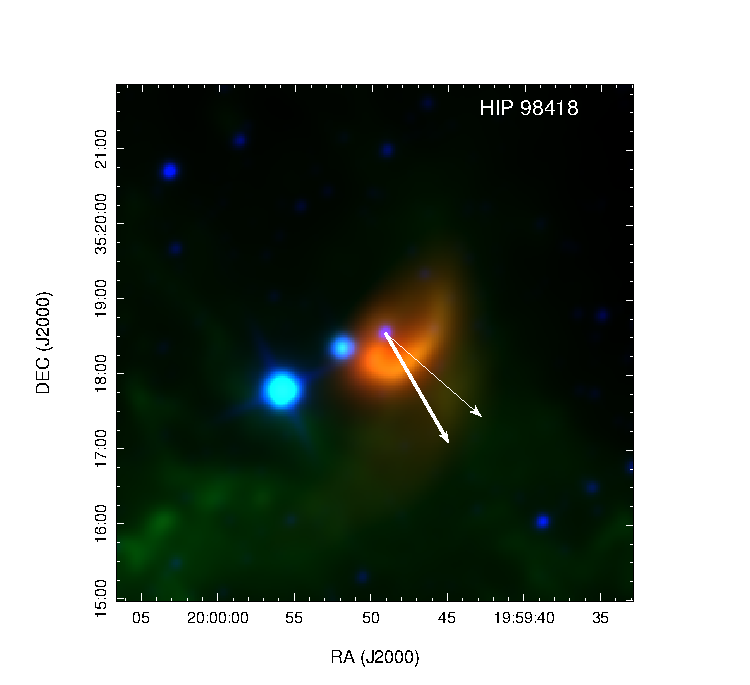}
\hfill
\includegraphics[width=0.455\textwidth]{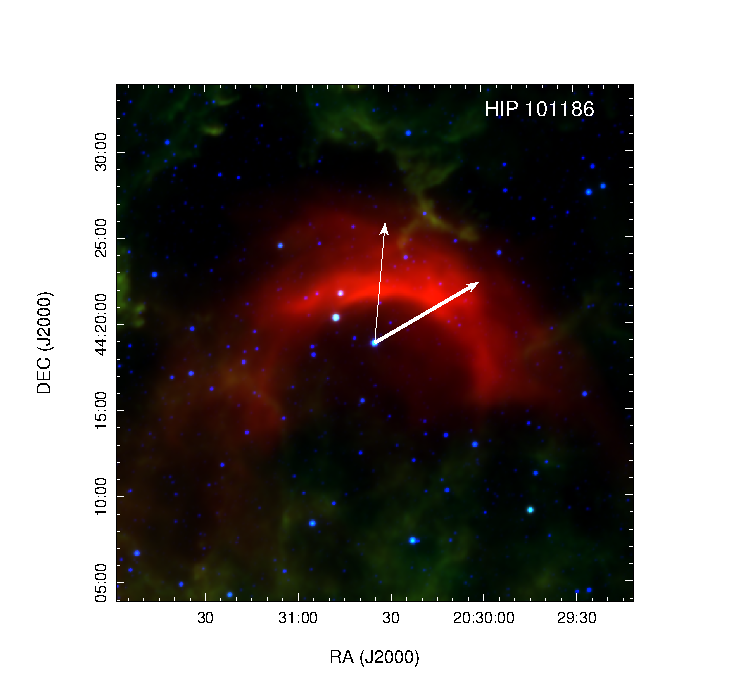}
\end{minipage} 

\caption{Group 3 bow-shock candidates WISE images. 
Red: band 4, $22.2 \, \mu$m. Green: band 3, $12.1 \, \mu$m. Blue: band 1, $3.4 \, \mu$m.
The label stands for the name in the final list of E-BOSS r2 BSCs. 
We have drawn $\mu_{\alpha} \cos \delta$ and $\mu_{\delta}$ to compose the total $\mu$.
The thick vectors represent the measured {\it Hipparcos} proper motions \citep{vanLeeuwen2007},
and the thinner vectors stand for the star proper motions but corrected
for the ISM motion caused by Galactic rotation. The vector lengths are not scaled with 
the original values. Compass as in HIP 44368 is the same for all RGB images.}
\label{BSC1}

\end{figure*} 

%-------------------------------------------------------------------

\begin{figure*}[t]

\begin{minipage}{\textwidth}
\centering
\includegraphics[width=0.465\textwidth]{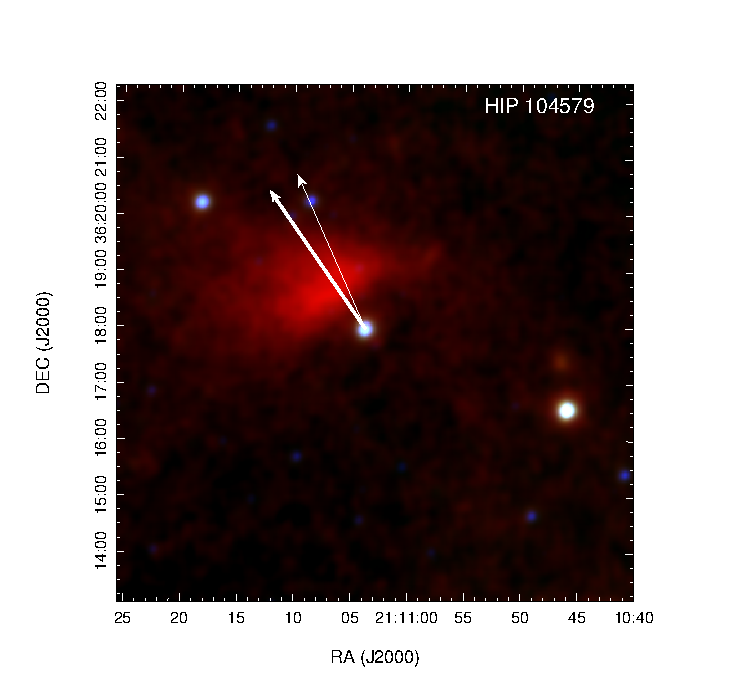}
\hfill
\includegraphics[width=0.465\textwidth]{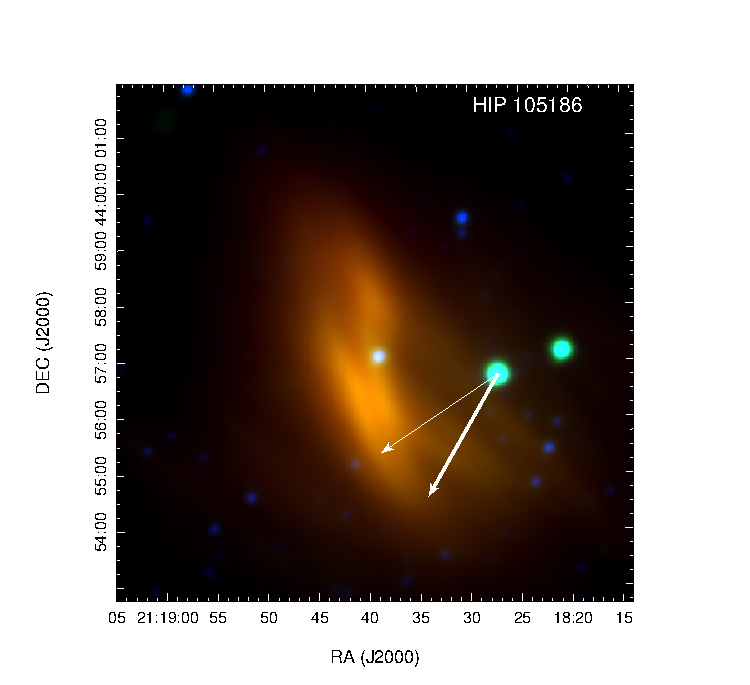}
\end{minipage} 

\begin{minipage}{\textwidth}
\centering
\includegraphics[width=0.465\textwidth]{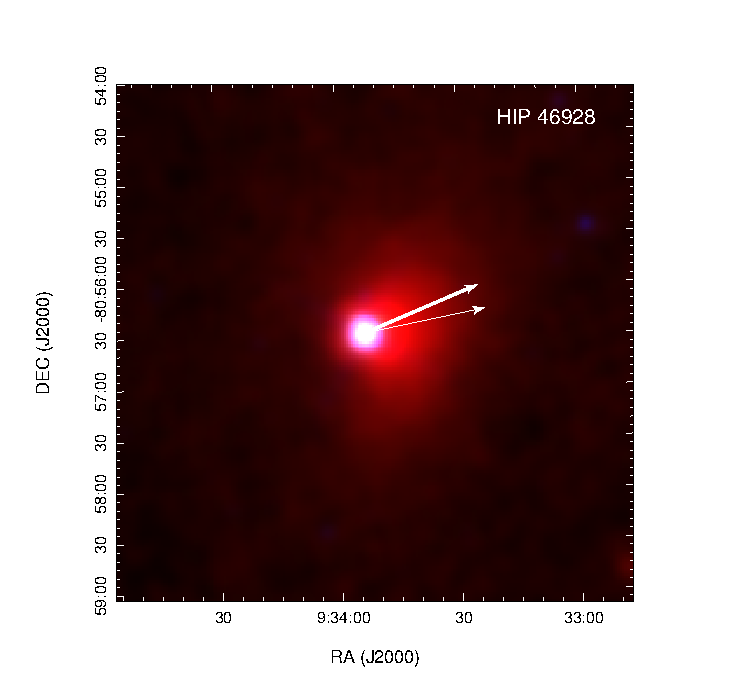}
\hfill
\includegraphics[width=0.465\textwidth]{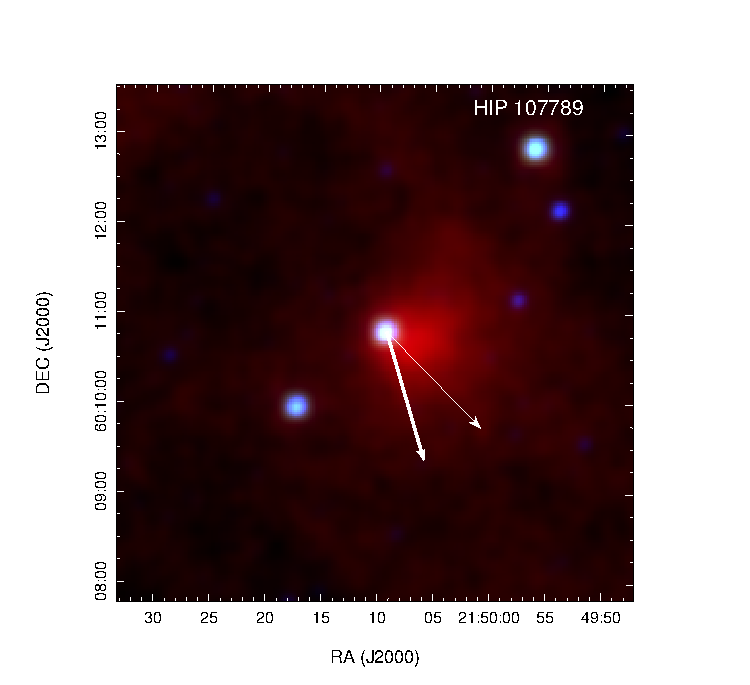}
\end{minipage} 

\begin{minipage}{\textwidth}
\centering
\includegraphics[width=0.465\textwidth]{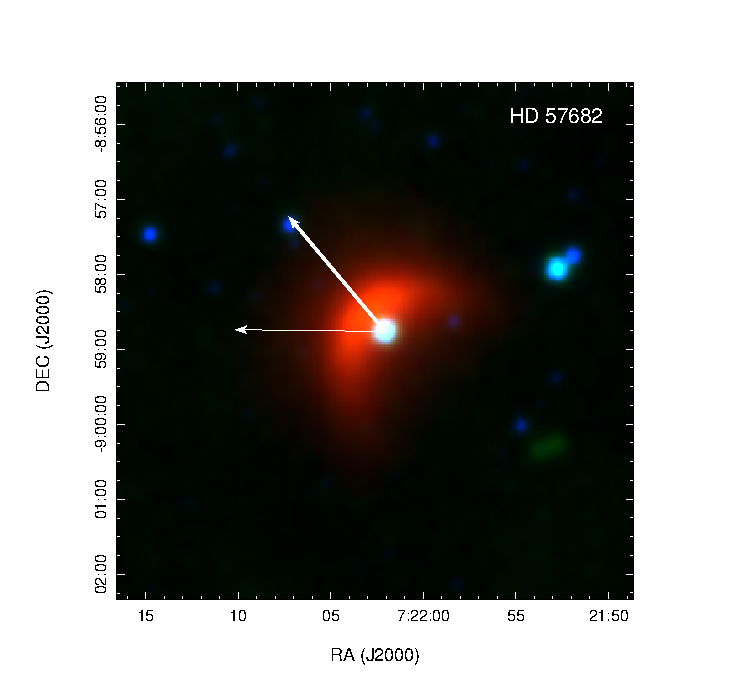}
\hfill
\includegraphics[width=0.465\textwidth]{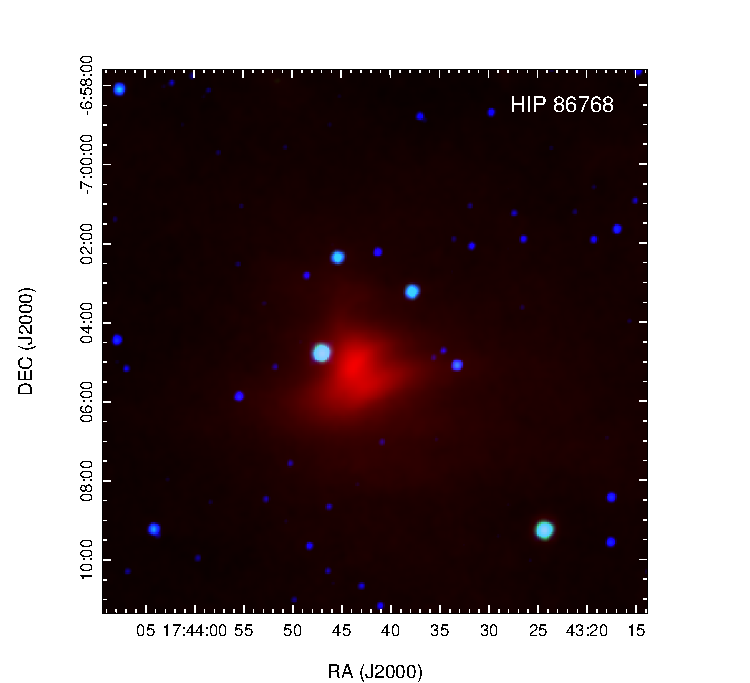}
\end{minipage} 

\caption{Groups 3, 4, 5, and 6 bow-shock candidates WISE images. 
Red: band 4, $22.2 \, \mu$m. Green: band 3, $12.1 \, \mu$m. Blue: band 1, $3.4 \, \mu$m.
The label stands for the name in the final list of E-BOSS r2 BSCs. 
We have drawn $\mu_{\alpha} \cos \delta$ and $\mu_{\delta}$ to compose 
the total $\mu$.
The thick vectors represent the measured {\it Hipparcos} proper motions \citep{vanLeeuwen2007},
and the thinner vectors stand for the star proper motions but corrected
for the ISM motion caused by Galactic rotation. The vector lengths are not scaled with 
the original values.}
\label{BSC2}

\end{figure*}

%-------------------------------------------------------------------

\begin{figure*}[t]

\begin{minipage}{\textwidth}
\centering
\includegraphics[width=0.465\textwidth]{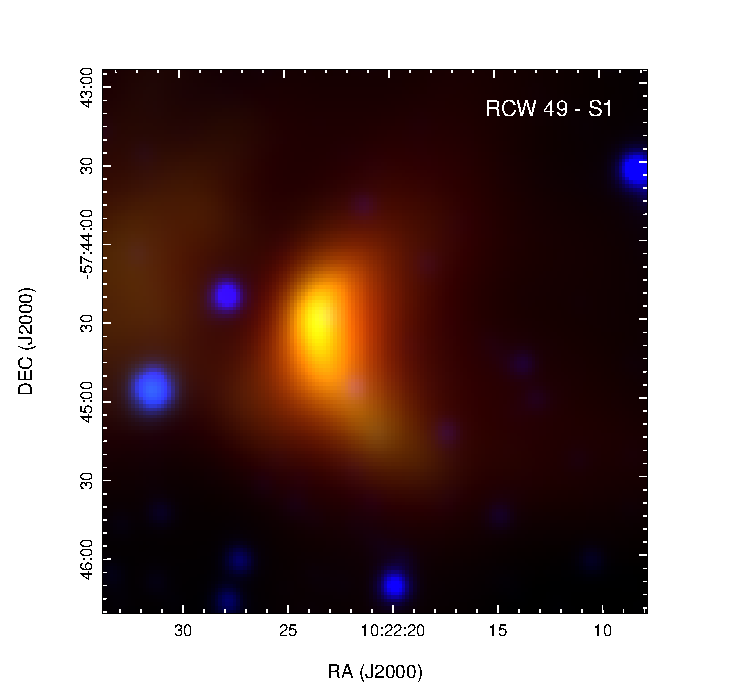}
\hfill
\includegraphics[width=0.465\textwidth]{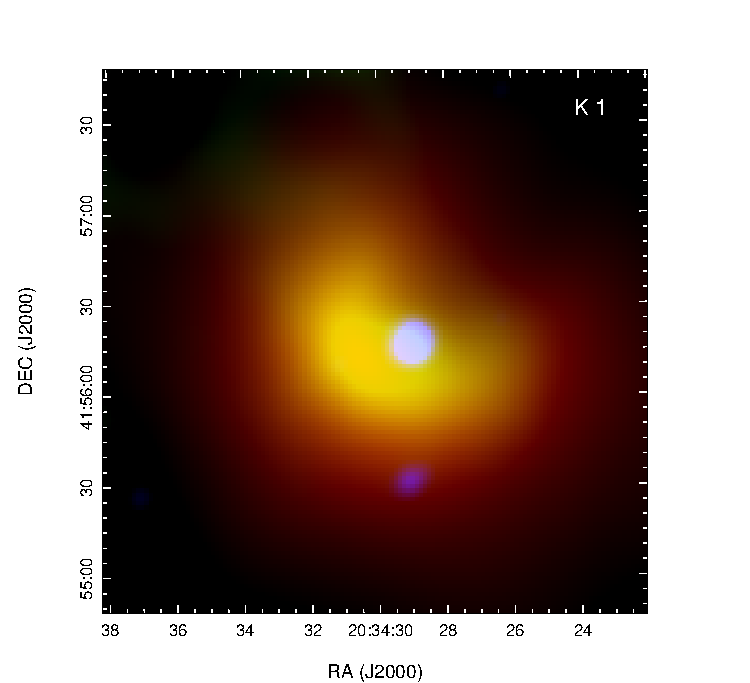}
\end{minipage} 

\begin{minipage}{\textwidth}
\centering
\includegraphics[width=0.465\textwidth]{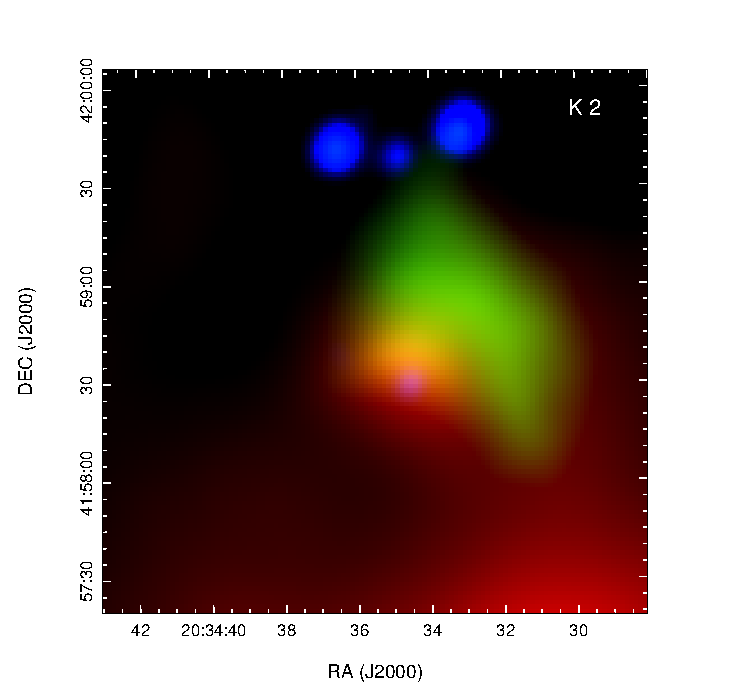}
\hfill
\includegraphics[width=0.465\textwidth]{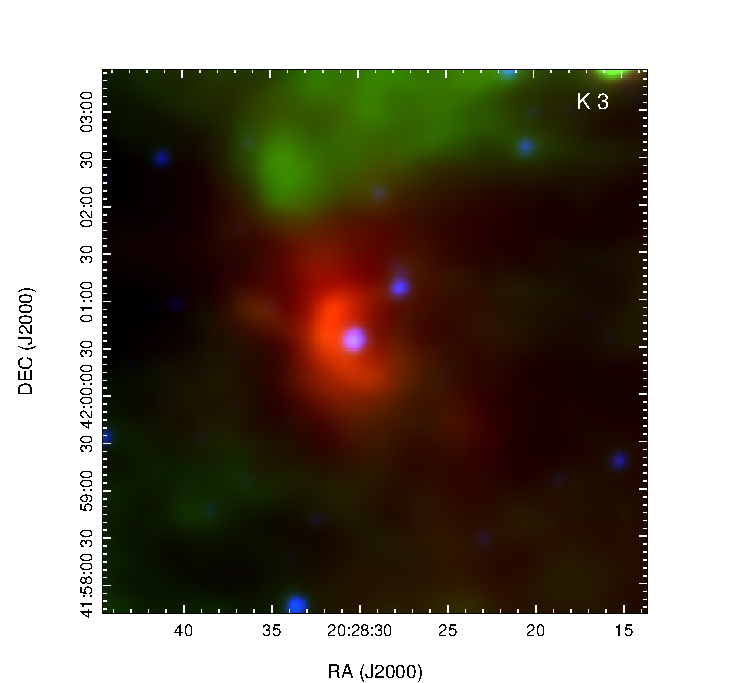}
\end{minipage} 

\begin{minipage}{\textwidth}
\centering
\includegraphics[width=0.465\textwidth]{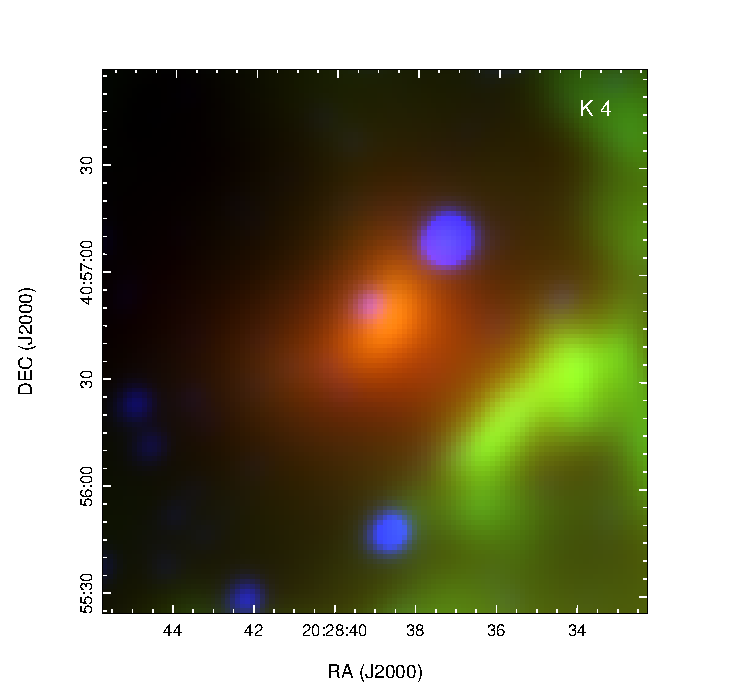}
\hfill
\includegraphics[width=0.465\textwidth]{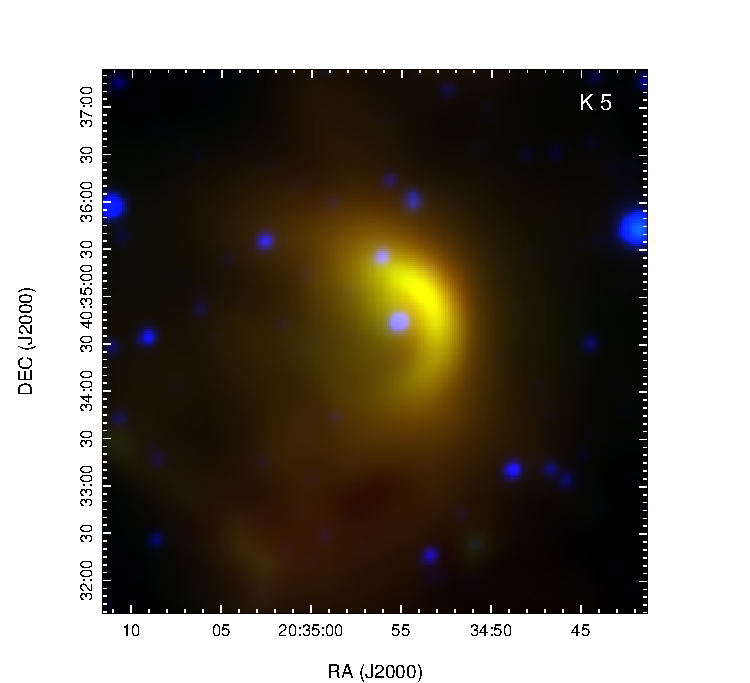}
\end{minipage} 

\caption{Group 7 bow-shock candidates WISE images. 
Red: band 4, $22.2 \, \mu$m. Green: band 3, $12.1 \, \mu$m. Blue: band 1, $3.4 \, \mu$m.
The label stands for the name in the final list of E-BOSS r2 BSCs.} 
\label{BSC3}

\end{figure*}

%-------------------------------------------------------------------

\begin{figure*}[t]

\begin{minipage}{\textwidth}
\centering
\includegraphics[width=0.465\textwidth]{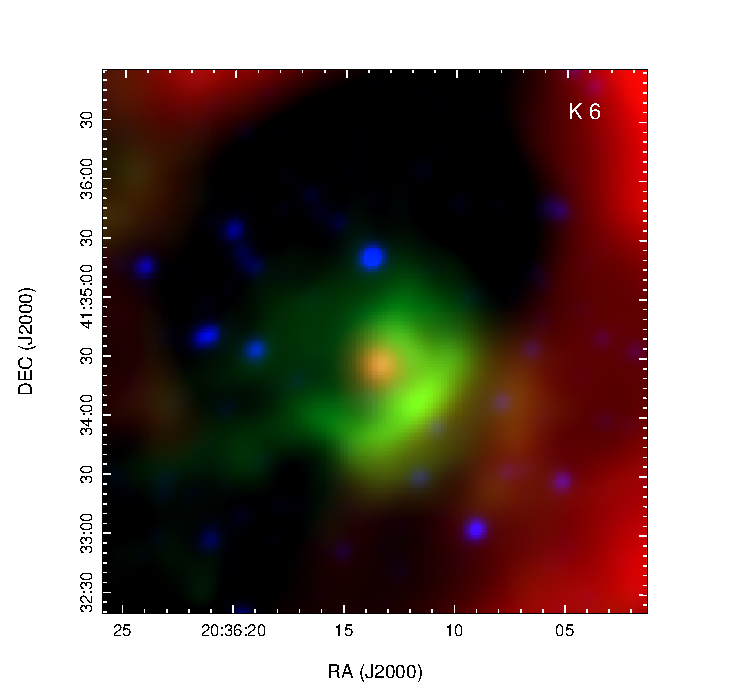}
\hfill
\includegraphics[width=0.465\textwidth]{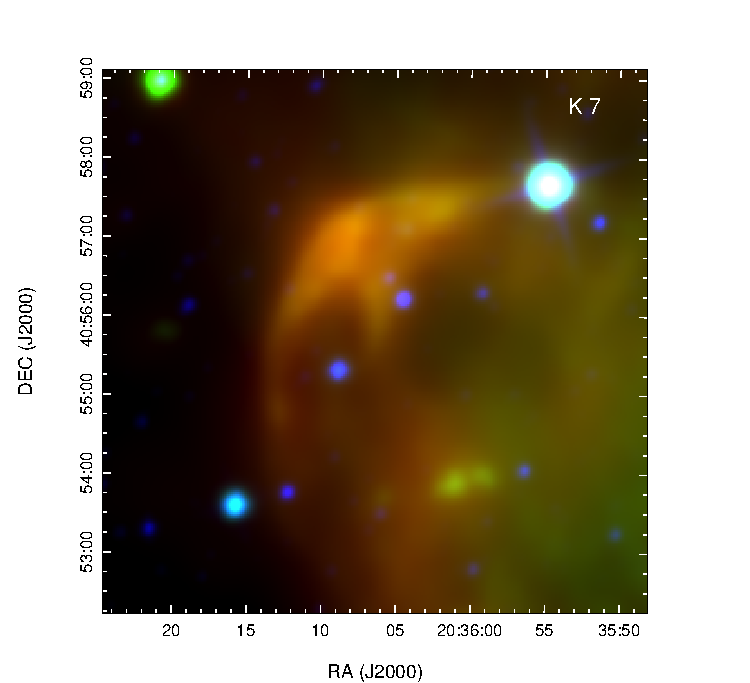}
\end{minipage} 

\begin{minipage}{\textwidth}
\centering
\includegraphics[width=0.465\textwidth]{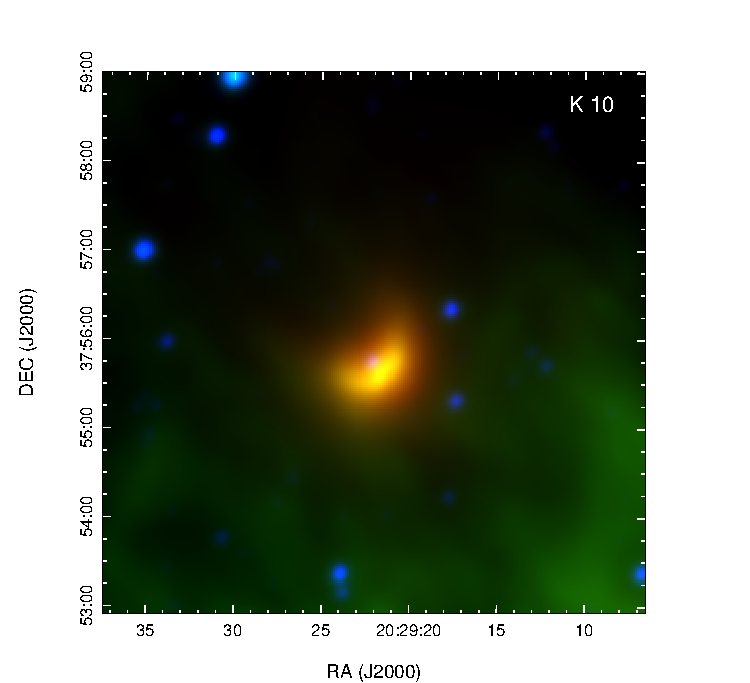}
\hfill
\includegraphics[width=0.465\textwidth]{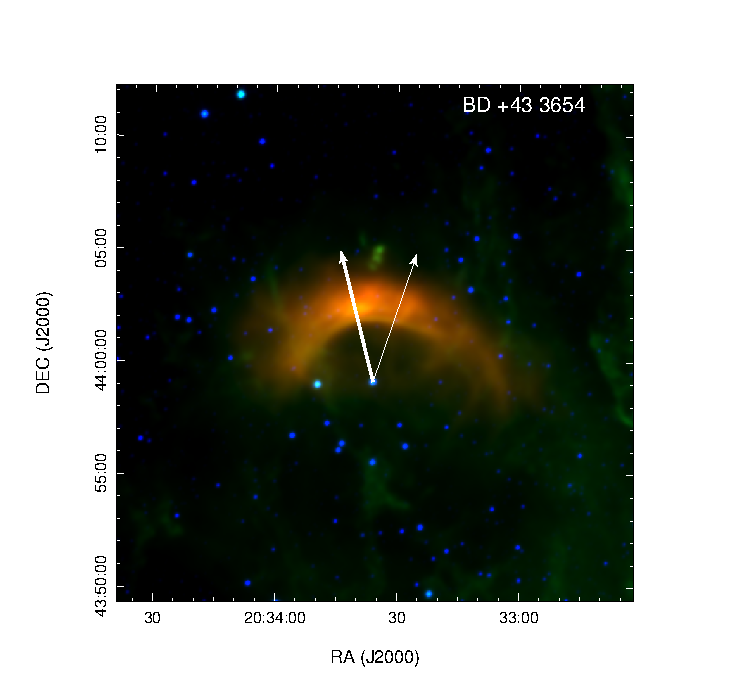}
\end{minipage} 

\begin{minipage}{\textwidth}
\centering
\includegraphics[width=0.465\textwidth]{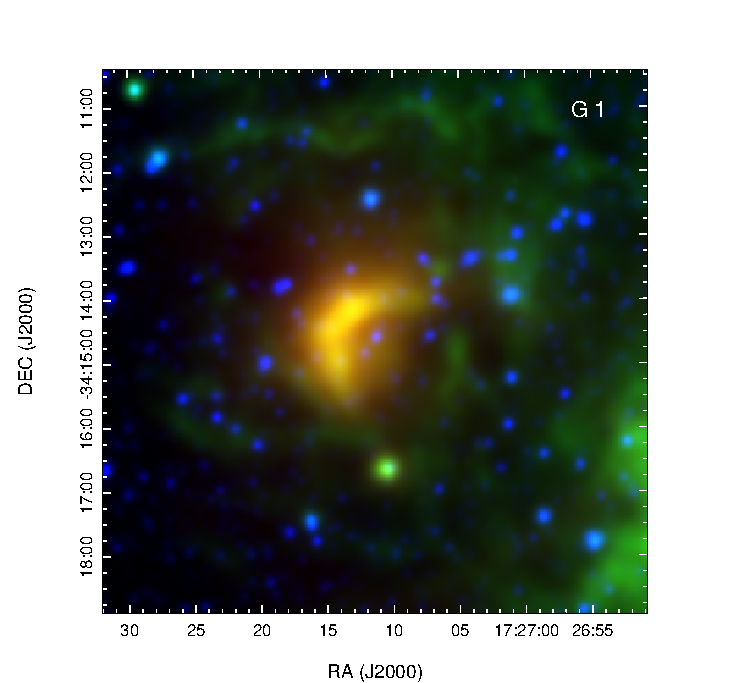}
\hfill
\includegraphics[width=0.465\textwidth]{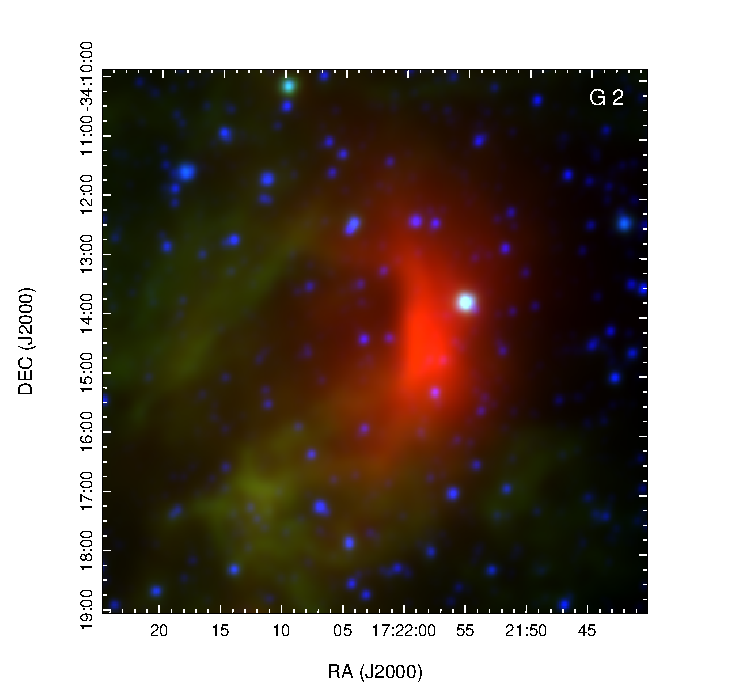}
\end{minipage} 

\caption{Group 7 bow-shock candidates WISE images. 
Red: band 4, $22.2 \, \mu$m. Green: band 3, $12.1 \, \mu$m. Blue: band 1, $3.4 \, \mu$m.
The label stands for the name in the final list of E-BOSS r2 BSCs.
For BD +43$^{\circ}$ 3654, we have drawn $\mu_{\alpha} \cos \delta$ and $\mu_{\delta}$ to compose 
the total $\mu$.
The thick vector represent the measured {\it Hipparcos} proper motions \citep{vanLeeuwen2007},
and the thinner vectors stand for the star proper motions but corrected
for the ISM motion caused by Galactic rotation. The vector lengths are not scaled with 
the original values.}
\label{BSC4}

\end{figure*}

%-------------------------------------------------------------------

\begin{figure*}[t]

\begin{minipage}{\textwidth}
\centering
\includegraphics[width=0.463\textwidth]{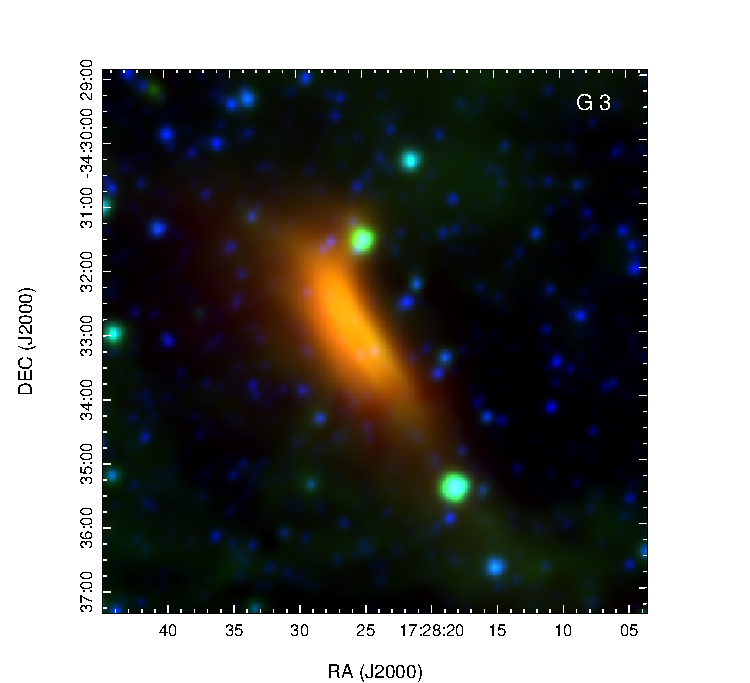}
\hfill
\includegraphics[width=0.463\textwidth]{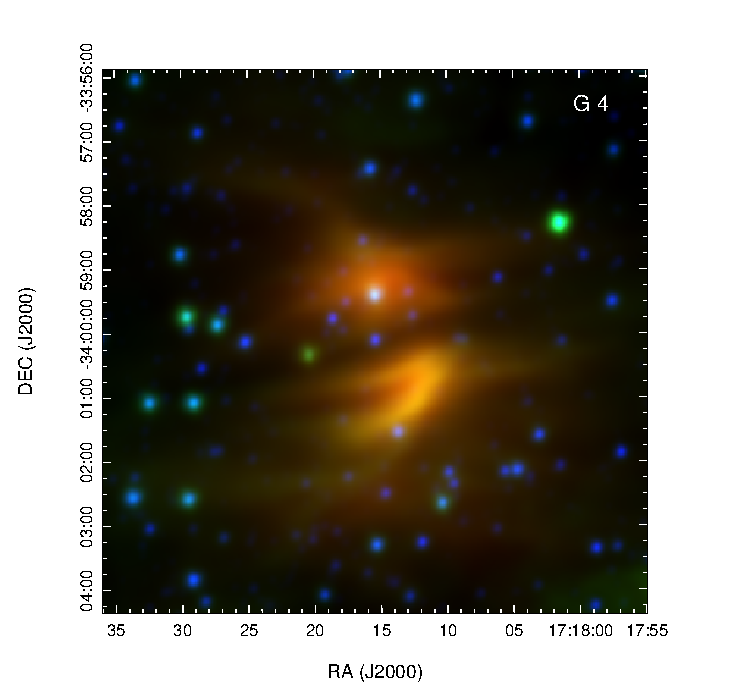}
\end{minipage} 

\begin{minipage}{\textwidth}
\centering
\includegraphics[width=0.463\textwidth]{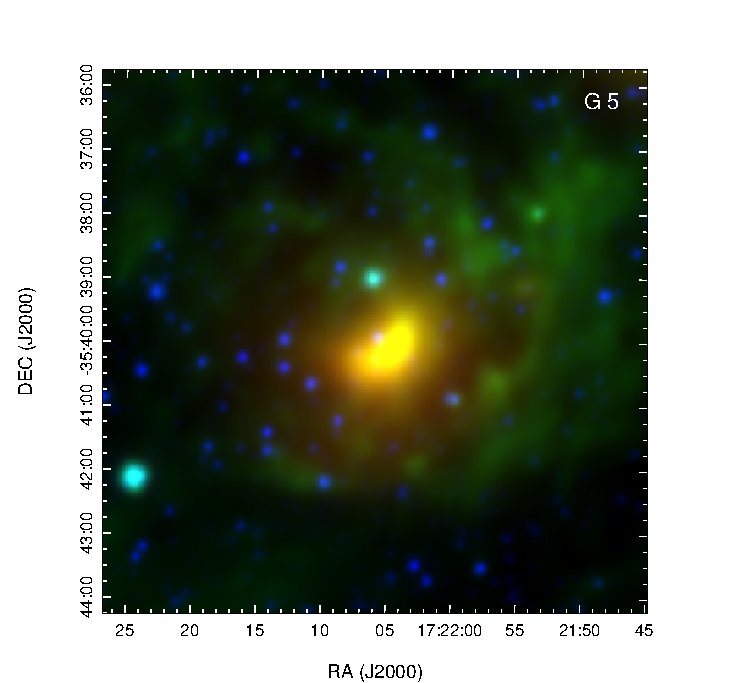}
\hfill
\includegraphics[width=0.463\textwidth]{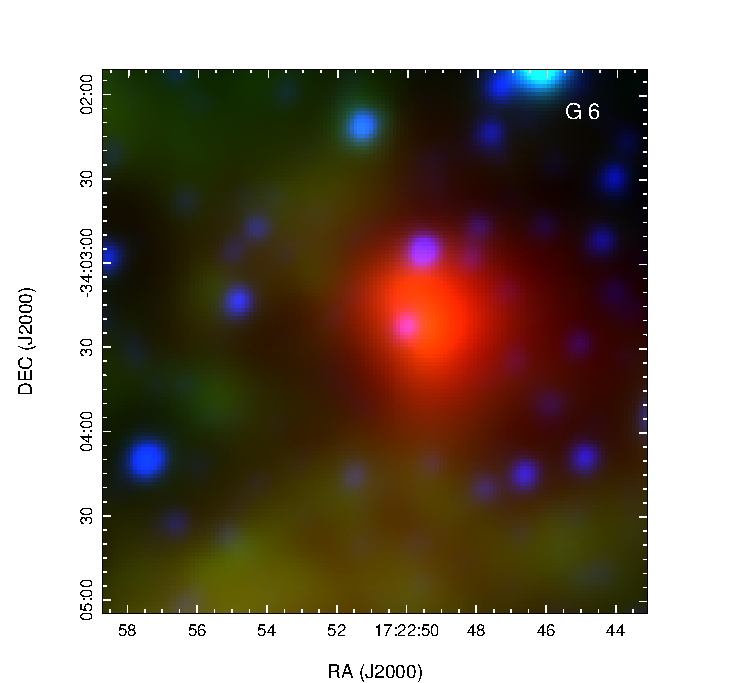}
\end{minipage} 

\begin{minipage}{\textwidth}
\centering
\includegraphics[width=0.463\textwidth]{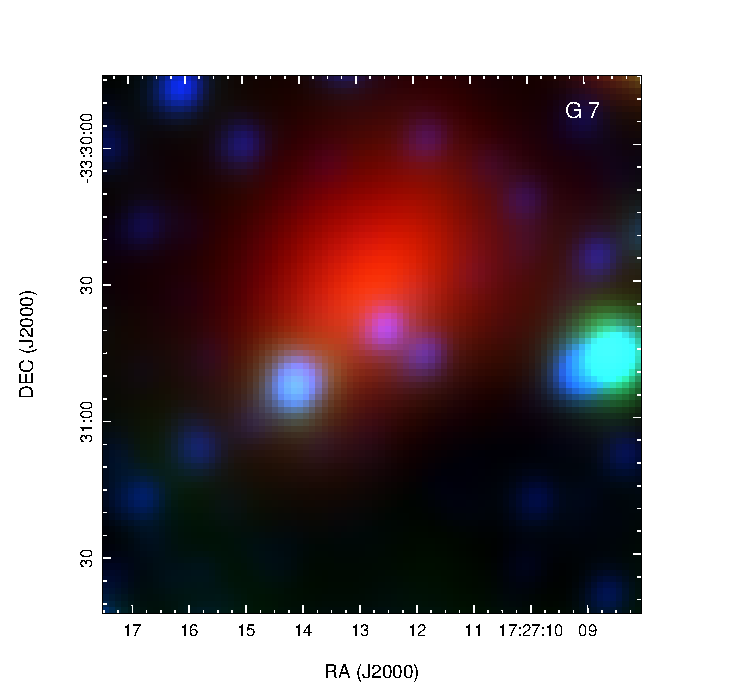}
\hfill
\includegraphics[width=0.463\textwidth]{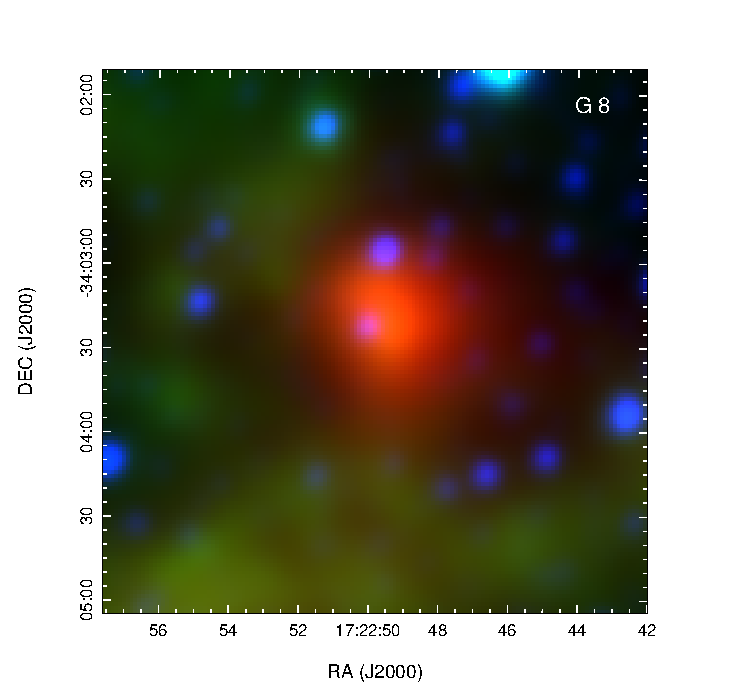}
\end{minipage} 

\caption{Group 7 bow-shock candidates WISE images. 
Red: band 4, $22.2 \, \mu$m. Green: band 3, $12.1 \, \mu$m. Blue: band 1, $3.4 \, \mu$m.
The label stands for the name in the final list of E-BOSS r2 BSCs. 
The BSC labeled G 4 has two structures; we have taken the 
BSC of the northern one because of the location of the star that might have generated it.}
\label{BSC5}

\end{figure*}

%-------------------------------------------------------------------

\begin{figure*}[t]

\begin{minipage}{\textwidth}
\centering
\includegraphics[width=0.47\textwidth]{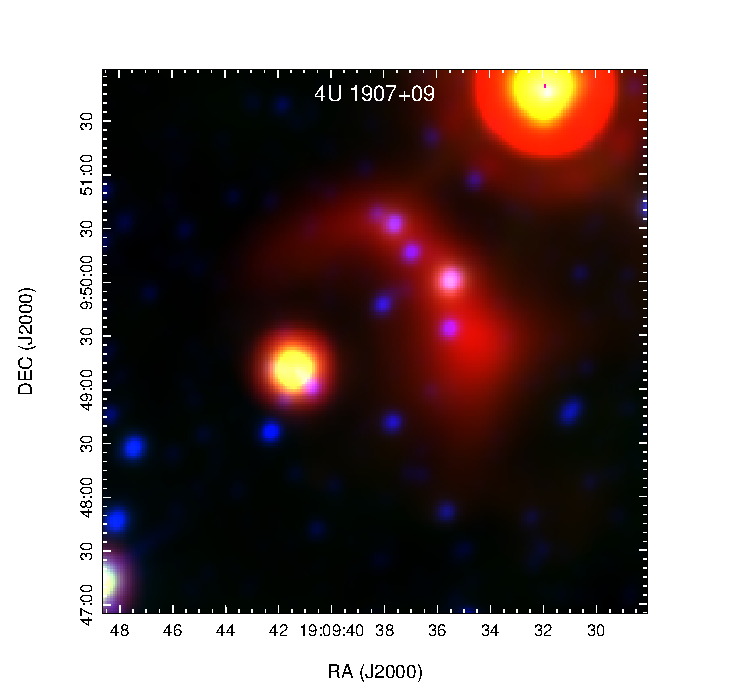}
\hfill
\includegraphics[width=0.47\textwidth]{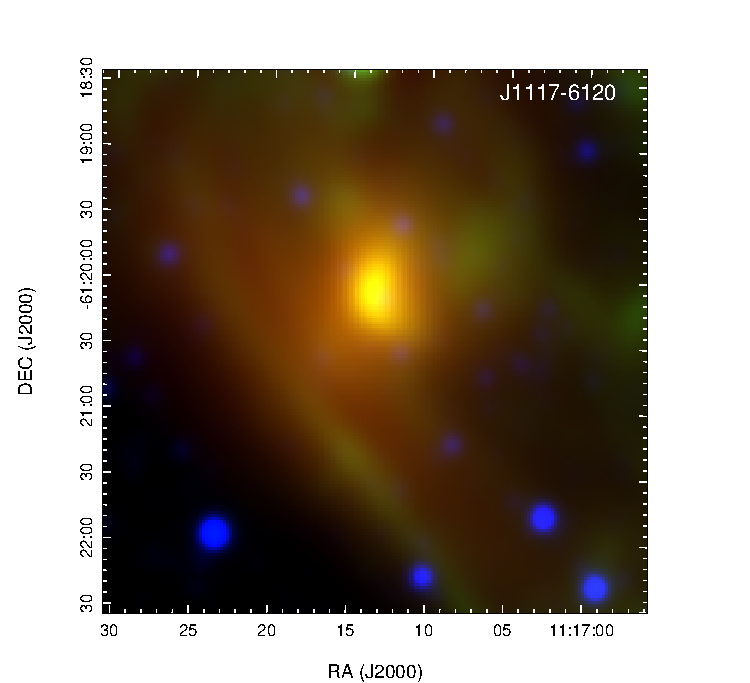}
\end{minipage} 

\begin{minipage}{\textwidth}
\centering
\includegraphics[width=0.465\textwidth]{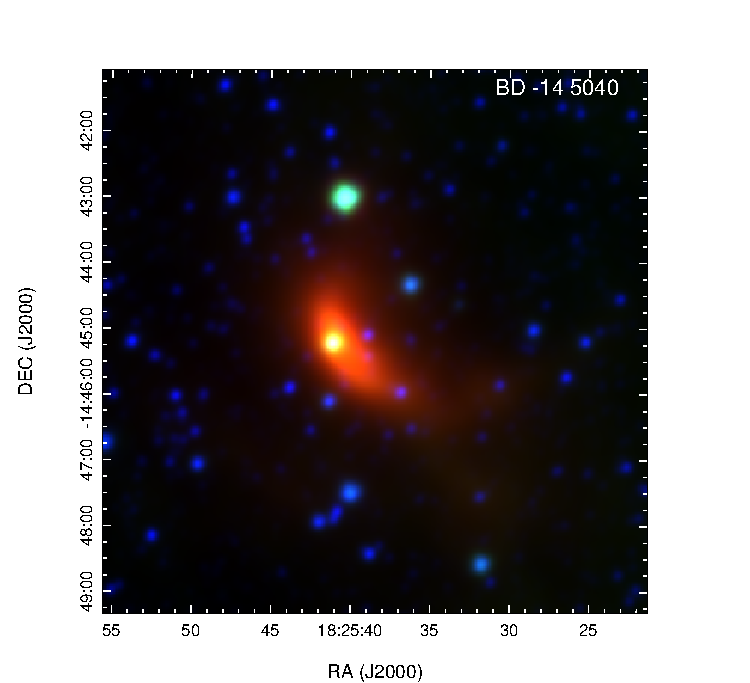}
\hfill
\includegraphics[width=0.465\textwidth]{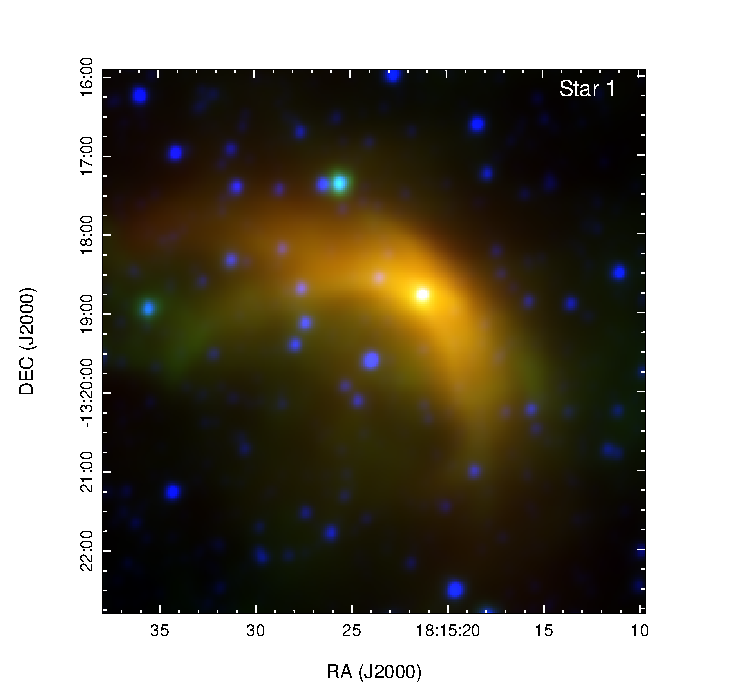}
\end{minipage} 

\begin{minipage}{\textwidth}
\centering
\includegraphics[width=0.465\textwidth]{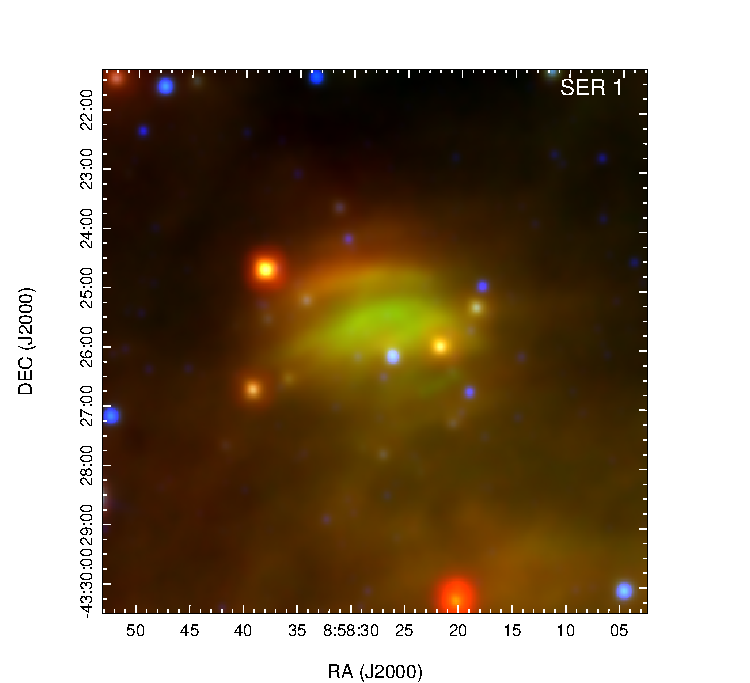}
\hfill
\includegraphics[width=0.465\textwidth]{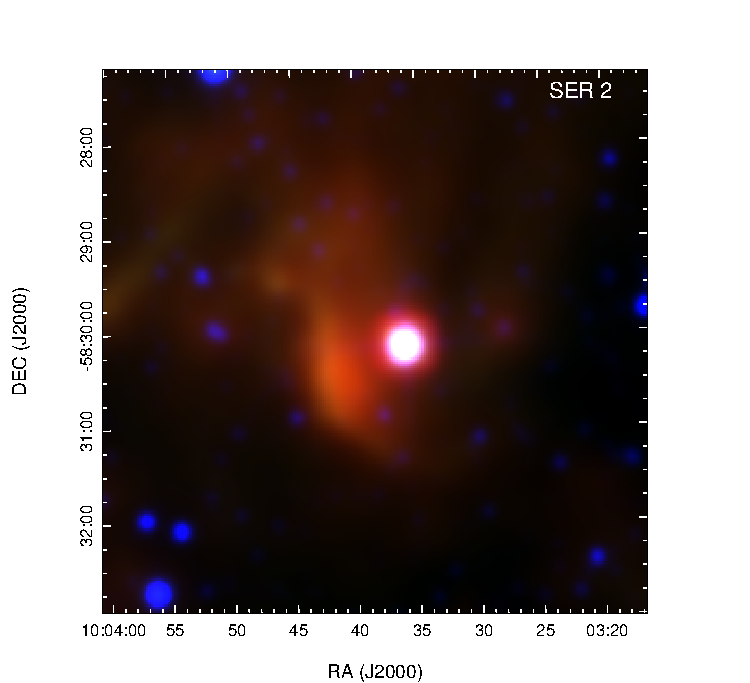}
\end{minipage} 

\caption{Group 7 bow-shock candidates WISE images. 
Red: band 4, $22.2 \, \mu$m. Green: band 3, $12.1 \, \mu$m. Blue: band 1, $3.4 \, \mu$m.
The label stands for the name in the final list of E-BOSS r2 BSCs.}
\label{BSC6}

\end{figure*}

%-------------------------------------------------------------------

\begin{figure*}[t]

\begin{minipage}{\textwidth}
\centering
\includegraphics[width=0.465\textwidth]{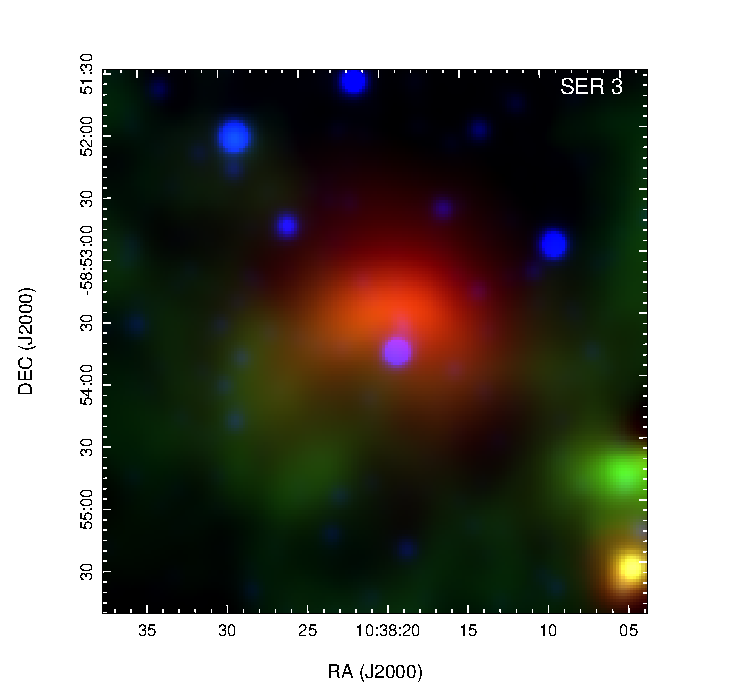}
\hfill
\includegraphics[width=0.465\textwidth]{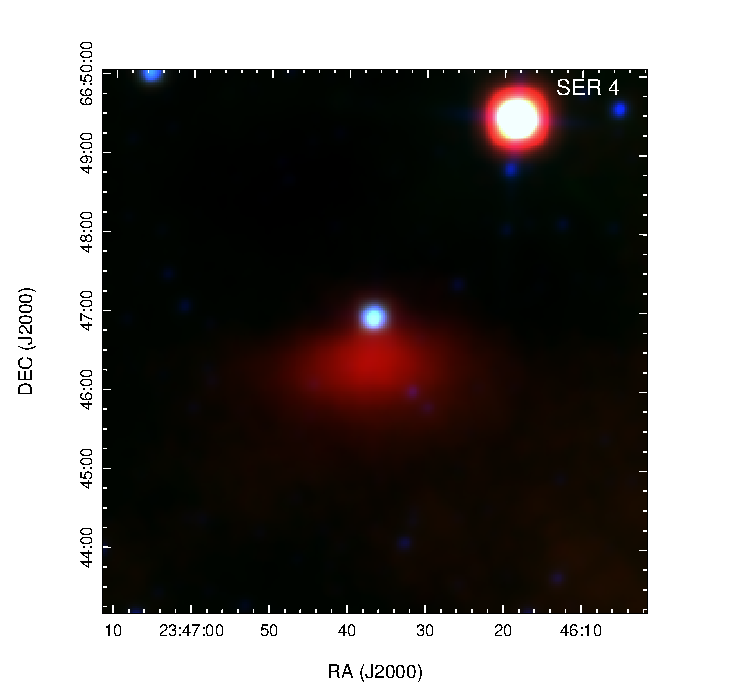}
\end{minipage}

\begin{minipage}{\textwidth}
\centering
\includegraphics[width=0.465\textwidth]{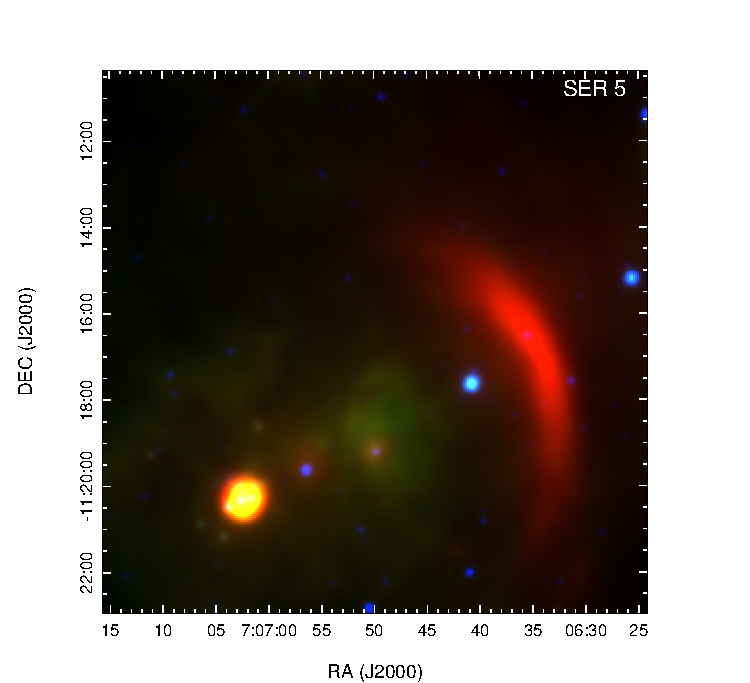}
\hfill
\includegraphics[width=0.465\textwidth]{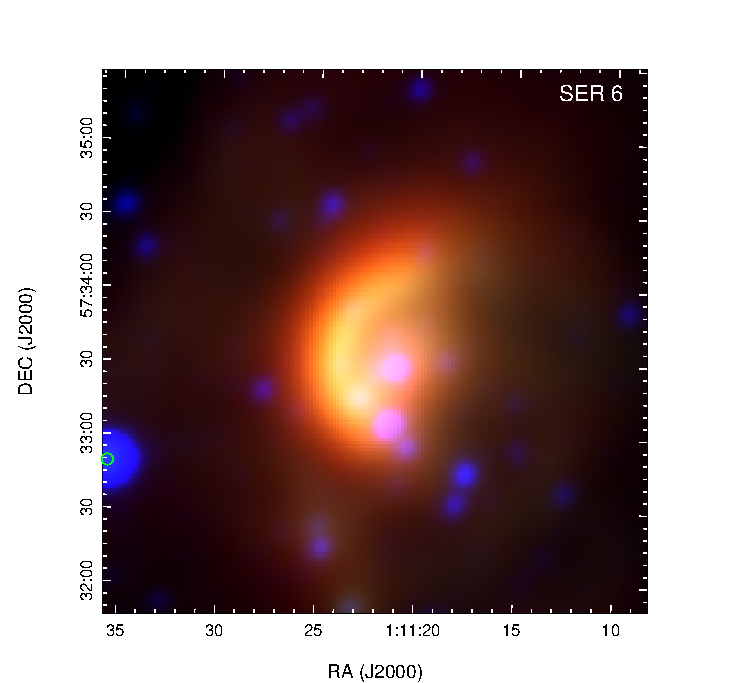}
\end{minipage}

\begin{minipage}{\textwidth}
\centering
\includegraphics[width=0.465\textwidth]{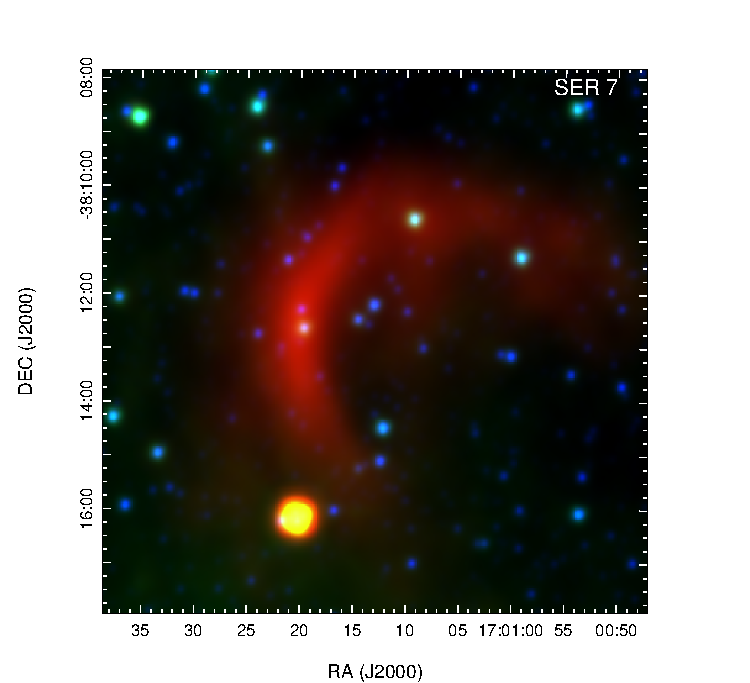}
\end{minipage} 

\caption{Group 7 bow-shock candidates WISE images. 
Red: band 4, $22.2 \, \mu$m. Green: band 3, $12.1 \, \mu$m. Blue: band 1, $3.4 \, \mu$m.
The label stands for the name in the final list of E-BOSS r2 BSCs.}
\label{BSC7}

\end{figure*}

%-------------------------------------------------------------------

\end{document}